\documentclass[aps,prd,preprint,floatfix,nofootinbib,longbibliography,a4paper]{revtex4-1}
\pdfoutput=1
\usepackage{color}
\usepackage{graphicx}
\usepackage{textcomp}
\usepackage{subfigure}
\usepackage{cancel}
\usepackage{feynmp}
\usepackage{amsmath,amssymb,array}
\usepackage{enumerate}
\usepackage{hhline}
\usepackage{cancel}
\newcommand{\mathsym}[1]{{}} 
\def\lsim{\:\raisebox{-1.1ex}{$\stackrel{\textstyle<}{\sim}$}\:}

\newcommand{\beqa}{\begin{eqnarray}}
\newcommand{\eeqa}{\end{eqnarray}}
\newcommand{\be}{\begin{equation}}
\newcommand{\ee}{\end{equation}}
\newcommand{\ba}{\begin{array}} 
\newcommand{\ea}{\end{array}}

\def\vev#1{\langle #1\rangle}

\begin{document} 
\vspace*{0.5cm}
\title{Weak scale right-handed neutrino as pseudo-Goldstone fermion of spontaneously broken $U(1)_{L_\mu-L_\tau}$}
\bigskip
\author{Anjan S. Joshipura}
\email{anjanjoshipura@gmail.com}
\author{Ketan M. Patel}
\email{kmpatel@prl.res.in}
\affiliation{Theoretical Physics Division, Physical Research Laboratory, Navarangpura, Ahmedabad-380 009, India.}

\begin{abstract}
Possibility of a Right-Handed (RH) neutrino being a Goldstone fermion of a spontaneously broken global $U(1)$ symmetry in a supersymmetric theory is considered. This fermion obtains mass from the supergravity effects leading to a RH neutrino at the electroweak scale with a mass similar to the gravitino mass. A prototype model realizing this scenario contains just three gauge singlet superfields needed for the type I seesaw mechanism. Masses of the other two neutrinos are determined by the $U(1)$ breaking scale which too can be around the electroweak scale. Light neutrinos obtain their masses in this scenario through (a) mixing with the RH neutrinos (type I seesaw), (b) mixing with neutralinos ($R$-parity breaking), (c) indirectly through mixing of the RH neutrinos with neutralinos, and (d) radiative corrections. All these contributions are described by the same set of a small number of underlying parameters and provide a very constrained and predictive framework for the neutrino masses which is investigated in detail for various choices of $U(1)$ symmetries. It is found that flavour independent $U(1)$ symmetries cannot describe neutrino masses if the soft supersymmetry breaking terms are flavour universal and one needs to consider flavour dependent symmetries. Considering a particular example of $L_\mu - L_\tau$ symmetry, it is shown that viable neutrino masses and mixing can be obtained without introducing any flavour violation in the soft sector. The leptonic couplings of Majoron are worked out in the model and shown to be consistent with various laboratory, astrophysical and cosmological constraints. The neutrino data allows sizeable couplings between the RH neutrinos and Higgsinos which can be used to probe the pseudo-Goldstone fermion at colliders through its displaced decay vertex.
\end{abstract}

\maketitle

\section{Introduction}
\label{sec:intro}

Extensions of the Standard Model (SM) by one or more Right-Handed (RH) singlet fermions provide the most popular mechanism for obtaining light neutrino masses \cite{Minkowski:1977sc,Yanagida:1979as,Mohapatra:1979ia,PhysRevD.22.2227}. The scale $M_N$ of the masses of these neutrinos is set by different underlying physics. $M_N$ can be close to the Planck scale in the absence of any symmetry protecting it. Grand unified theories based on $SO(10)$ group can lead to $M_N$ in the range $10^{10}-10^{16}$ GeV \cite{Fritzsch:1974nn,GellMann:1980vs}. Smaller $M_N$ close to TeV scale can be linked to theories with a low scale gauge symmetry breaking  such as $U(1)_{B-L}$ or the left right symmetry. The supersymmetric theories with a spontaneously broken global symmetry provide an alternate mechanism for protecting the RH neutrino masses upto TeV scale or less \cite{Chun:1995js,Chun:1995bb}.  If the global symmetry is broken by singlet superfield(s) and the supersymmetry remains unbroken in the process then the fermionic partner of the Goldstone Boson - the Goldstone Fermion - would also remain massless. Once the supersymmetry is broken, say by supergravity effects, this fermion can pick up a mass around the gravitino mass $m_{3/2}\sim$ TeV or less. This pseudo-Goldstone Fermion (pGF) can be identified with one of the sterile neutrinos. Implementation of this mechanism generically requires more than one singlet superfields which may be identified with the additional RH neutrinos participating in the seesaw mechanism.

The minimal choice of the global symmetry is $U(1)$. The basic assumption is that the underlying theory is supersymmetric and invariant under $U(1)$. The $U(1)$ is assumed to break spontaneously keeping supersymmetry intact. A prototype model satisfying these requirements contains just three singlet chiral superfields $\hat{N}_\alpha\equiv(\hat{Y},\hat{N},\hat{N}')$ with $U(1)$ charges $(0,-1,1)$ and the superpotential
\be \label{WN0}
W_N = \lambda\,(\hat{N} \hat{N}^\prime - f^2)\, \hat{Y}\,. \ee
As usual, a chiral superfield $\hat{N}_\alpha$ contains a complex scalar $\tilde{N}_\alpha$ and its fermionic partner which we denote as $N_\alpha$. We identify the fermionic parts of $\hat{N}_\alpha$ with the three RH neutrinos. Then Eq. (\ref{WN0}), apart from breaking $U(1)$ also determines the zeroeth order structure of the RH neutrino mass matrix which gets corrected by the supergravity effects. The Dirac mass terms required to complete the seesaw mechanism arise from the following superpotetial
\be \label{WD}
W_D=\lambda_{\alpha\beta}\, \hat{L}_\alpha \hat{N}_\beta \hat{H}_2\,.\ee
The structure of  $\lambda_{\alpha\beta} $ is determined by the choice of the quantum numbers of $(L_\alpha,H_2)$ under $U(1)$ symmetry and we will consider various choices.

Eqs. (\ref{WN0}) and (\ref{WD}) contribute to neutrino masses in two different ways. One obtains non-zero Vacuum Expectation Values (VEVs) for scalars of $\hat{N}$, $\hat{N}^\prime$ at the global supersymmetric minimum from Eq. (\ref{WN0}). These VEVs  generate the masses for two  of the RH neutrinos. The third RH neutrino obtains its mass from the supersymmetry  breaking effects and the three neutrinos  together generate the seesaw masses for the light neutrinos. VEVs for $\hat{N}$, $\hat{N}^\prime$ also lead to spontaneous bilinear $R$-parity violation through Eq. (\ref{WD}) which leads to additional contribution to neutrino masses through their mixing with neutralinos (see for example \cite{Barbier:2004ez} and references therein). These two entirely different contributions to neutrino masses are governed by the same set of parameters appearing in Eqs. (\ref{WN0},\ref{WD}) and provide a very constrained and predictive framework for neutrino masses. Moreover, a simple requirement that the $R$-parity violating  contribution to neutrino masses is smaller or of the order of the neutrino masses lead to a an upper bound on the RH sneutrino VEVs since the  bilinear $R$-parity breaking parameters are proportional to these VEVs. This bound is around TeV but is model dependent as the the standard seesaw also  contributes to the neutrino masses. Since the RH sneutrino VEVs control the masses of the other two RH neutrinos, they also have masses around TeV. As a result, the masses of all three RH neutrinos are around TeV or less and one obtains a viable scenario for the TeV scale RH masses  without invoking any additional gauge symmetry, see \cite{Deppisch:2015qwa} for a review on weak scale RH neutrinos.

A phenomenological motivation to consider seesaw models with TeV scale RH neutrinos is the possibility of producing and detecting them at colliders. Disadvantage of these models on the other hand is the smallness of the active sterile mixing required to obtain the correct light neutrino masses. Since this mixing governs the production and decay of the RH neutrino in the minimal seesaw model it is not possible to observe such RH neutrino within the minimum set up\footnote{There exist special textures of the Dirac and Majorana neutrino mass matrices which lead to small neutrino masses and large active sterile mixing even within the minimal seesaw, see for example \cite{Kersten:2007vk,Lee:2013htl,Chattopadhyay:2017zvs}.}. This changes if one or more RH neutrinos are pGF of the type considered above. In this case, the symmetry breaking VEV of the RH sneutrino directly couples the pGF to Higgsinos and their production does not require suppressed mixing  as is the case with the standard seesaw model. Relatively small mixing $|V_{N\nu_i}| \sim 10^{-6}$ of the pGF with neutrinos make them long lived and pGF can lead to displaced vertex signature at in the collider experiments. This possibility has been studied extensively recently \cite{Lavignac:2020yld} and can also be realized within the scenario presented in this paper.

We introduce the basic framework in the next section choosing the above mentioned case of the spontaneously broken $U(1)$ symmetry. We then discuss structure of neutrino masses in section \ref{sec:FIU1} assuming $U(1)$ to be flavour independent. It is shown that models in this category cannot reproduce the neutrino masses correctly if the soft supersymmetry (SUSY) breaking terms are assumed flavour universal. We then discuss the case of the $U(1)_{L_\mu-L_\tau}$ symmetry in section \ref{sec:lmultau} which allows correct description of the observed neutrino spectrum even with the universal soft terms. The framework implies an existence of Goldstone Boson (GB) of sponteneously broken $U(1)_{L_\mu-L_\tau}$ symmetry, namely the Majoron \cite{Chikashige:1980ui}. In section \ref{sec:Majoron}, we derive the couplings of Majoron with the charged and neutral leptons and discuss their phenomenological implications. We also discuss about the possibility to detect the lightest RH neutrino in the direct search experiment in section \ref{sec:direct}. The work is summarized in section \ref{sec:summary}.

\section{Basic framework}
\label{sec:BF}
The starting point of the models we discuss is the renormalizable superpotential given in Eq. (\ref{WN0}). The other terms in $W_N$, such as $Y^2$ or $Y^3$, can be forbidden by an additional symmetry like $R$ symmetry. The minimization of the potential $V_N$ obtained from $W_N$ leads to a global SUSY preserving minimum corresponding to 
\be \label{vev}
\langle \tilde{N} \rangle = U s_\phi\,,~~\langle \tilde{N}^\prime \rangle = U c_\phi\,,~~\langle \tilde{Y} \rangle = 0\,,\ee
where $s_\phi = \sin\phi$, $c_\phi = \cos\phi$ and 
\be \label{Uf}
U = \frac{f}{\sqrt{s_\phi c_\phi}}\,,\ee
represents $U(1)$ breaking scale. Even in the presence of terms like $Y^2$ and  $Y^3$ in $W_N$, Eq. (\ref{vev}) describes the global minimum of the potential.

The spontaneous breaking of global $U(1)$ symmetry implies a massless Goldstone boson. Writing the scalar fields as 
\be \label{repl_N}
\tilde{N}_\alpha = \langle \tilde{N}_\alpha \rangle + \frac{1}{\sqrt{2}}\left( \tilde{N}_{R \alpha} + i\tilde{N}_{I \alpha} \right)\,, \ee
at the minimum of the potential, we obtain the Goldstone boson as
\be \label{GB}
G_B = c_\phi \tilde{N}^\prime_{I} - s_\phi \tilde{N}_{I}\,.\ee
Its fermionic partner also remains massless in the SUSY preserving limit and it is given by a similar linear combination
\be \label{GF}
G_F = c_\phi N^\prime - s_\phi N\,.\ee
This can also be seen explicitly by constructing singlet fermion mass matrix from Eq. (\ref{WN0}). It is obtained as
\be \label{MN0}
M_N^0=\lambda U \left(\ba{ccc}
0 & c_\phi &  s_\phi\\
 c_\phi & 0 & 0\\
 s_\phi & 0 & 0\\ \ea \right)\,. \ee
The diagonalization of the above matrix leads to a massless state given by Eq. (\ref{GF}) and two massive and degenerate states with masses $|\lambda U|$.

The fermion mass matrix in Eq. (\ref{MN0}) changes in the presence of the local SUSY breaking terms. These terms shift the VEVs of $\tilde{N},\tilde{N}'$ by ${\cal O}(m_{3/2})$ and also  generate a nonzero VEV for $\tilde{Y}$ field \cite{Goto:1991gq}. The later makes the Goldstone fermion massive. These shifts follow from the minimisation of the potential with the soft SUSY terms \cite{Chun:1992zk,Chun:1995hc}. If the kinetic terms in Kahler potential $K$ are minimal and hidden sector is assumed to have Polonyi form then the shifts are given by \cite{Lahanas:1986uc}
\be \label{yvev}
\delta \langle \tilde{N}\rangle = \delta \langle \tilde{N}' \rangle = \frac{m_{3/2}}{\lambda} \left(1+\frac{f^2\lambda^2}{m_{3/2}^2} \right)^{\frac{1}{2}}\,,~~~ \delta \langle \tilde{Y} \rangle = \frac{m_{3/2}}{\lambda}\,.\ee
Shifts for  general soft terms are calculated in \cite{Chun:1995hc}. The fermion mass matrix in SUGRA is given  by
\be \label{fermasses}
M_{ab}=M_P\, e^{\frac{G}{2M_P^2}}\, \left(G_{ab}+\frac{1}{3} \frac{G_a G_b}{M_P^2} -G_c\,(G^{-1})^c_d\, G^d_{ab} \right)\,.\ee
with 
\be \label{G}
G=K+M_P^2~\ln|W/M_P^3|^2\,,\ee
$G_a = \frac{\delta G}{\delta \phi^a}$, $\phi^a=\phi_a^*$ is antichiral superfield and
\be \label{wsusy}
W = g(z) + W_N (N_\alpha)\,.\ee
Here, $g(z)$ describes the hidden sector potential and $W_N$ is given by Eq. (\ref{WN0}). We evaluate the fermion mass matrix by  substituting $\vev{z}=b M_P$, $\vev{W}\sim\vev{g(z)} \sim \frac{m_{3/2} M_P^2}{c}$ where $c = e^{b^2/2}$ is a numerical constant and remembering that now  $\lambda \langle \tilde{Y} \rangle  \approx {\cal O}(m_{3/2})$. The gravitino mass is defined as $m_{3/2}=M_P\, \exp(\langle G \rangle / 2M_P^2)$. The fermion mass matrix following from Eq. (\ref{fermasses}) is then given by
\be \label{local}
M_N^{\rm loc} = c\, m_{3/2} \left(\ba{ccc}
{\cal O} \left(\frac{U^2}{M_P^2}\right) & \frac{\lambda U c_\phi}{m_{3/2}} &\frac{\lambda U s_\phi}{m_{3/2}}\\
\frac{\lambda U c_\phi}{m_{3/2}} & {\cal O} \left(\frac{U^2}{M_P^2}\right) &  \frac{\lambda \vev{\hat{Y}} }{m_{3/2}}\\
\frac{\lambda U s_\phi}{m_{3/2}} & \frac{\lambda \vev{\hat{Y}} }{m_{3/2}} & {\cal O} \left(\frac{U^2}{M_P^2}\right) \\ \ea \right)\,.\ee
The above mass matrix approximately coincides with the $M_N^0$ in Eq.  (\ref{MN0}) if one effectively assumes nonzero VEV for $\tilde{Y}$ in the latter. The other corrections induced by SUSY breaking are ${\cal O}(U^2/M_P^2)$ and they are very small. Therefore, we approximate the singlet fermion mass matrix in the presence of local SUSY breaking as
\be \label{MN}
M_N \simeq \lambda U \left(\ba{ccc}
0 & c_\phi &  s_\phi\\
 c_\phi & 0 & \epsilon\\
 s_\phi & \epsilon & 0\\ \ea \right)\,, \ee
where $\epsilon = \vev{\hat{Y}}/U < 1$ and we have absorbed $c$ in redefining $\lambda$. Note that $U$ denotes now a shifted VEV and it does not strictly satisfy Eq. (\ref{Uf}). The SUSY breaking generates the mass for the lightest RH neutrino and split the masses of the other two singlet neutrinos by ${\cal O}(m_{3/2})$. Explicitly,
\be \label{masses}
M_{N_1} \approx  |2 \epsilon \lambda U  c_\phi s_\phi|\,,~~ M_{N_{2,3}} \approx |\lambda U (1 \pm \epsilon c_\phi s_\phi)|\,.\ee
The lightest RH neutrino is now a pseudo-Goldstone fermion which can be written as\be \label{pGF}
N_1 \simeq \cos\phi\, N^\prime - \sin\phi\, N -  \epsilon \cos 2\phi\, Y\,,\ee
at the leading order in $\epsilon$. 

\section{Flavour independent $U(1)$ symmetries and neutrino masses}
\label{sec:FIU1}
We now discuss the interaction of the RH neutrinos with the SM leptons, Higgs and their superpartners.
This depends on the choice of the symmetry $U(1)$. There are four possible choices. $U(1)$ is (a) a ``hidden" symmetry
acting only on the RH fields $N_\alpha$, or (b) a leptonic symmetry acting on $L_\alpha$ and $N_\alpha$, or (c) a Peccei Quinn like symmetry acting on the Higgs doublet(s), on some of the right-handed quarks and may also be on the leptonic fields, or (d) a leptonic family symmetry acting only on leptons and distinguishing among them. We discuss the case (d) in detail in the next section. The cases (a-c) can all be described by a general superpotential 
\be \label{Wgen}
W = \lambda_\alpha\, \hat{L}_\alpha \hat{H}_2 \hat{S}_1\,+\,\kappa \, \hat{H}_1 \hat{H}_2\hat{S_2}\,
\ee
where we have only shown the part relevant for the neutrino masses.
Here $\hat{S}_{1,2}$ can be any of the three RH fields $\hat{N}_\alpha$ depending on the choice of the $U(1)$ symmetry.
For the case (a), one has $\hat{S}_1=\hat{S}_2=\hat{Y}$. The case (b) corresponds to $\kappa=0$ and $\hat{S}_1= \hat{N}$ if $L_\alpha$ have $U(1)$ charge is +1. $W$ in this case contains an additional $\mu$ term. The choice of $\hat{S}_{1,2}$ in case (c) depends on the $U(1)$ charges of $\hat{H}_{1,2}$.  

The cases (a-c) are similar and we consider the case (c) with $\hat{S}_1=\hat{N}$ and $\hat{S}_2=\hat{N}'$ for definiteness. Without loss of generality, the parameters $\lambda_\alpha$ and $\kappa$ can be chosen real.
We shift the fields $\hat{N},\hat{N}'$ by the VEV of their scalar components. This leads to the following superpotetial
\be \label{Wgenshifted}
W = \lambda_\alpha\, \hat{L}_\alpha \hat{H}_2 \hat{N}\,+\,\kappa \, \hat{H}_1 \hat{H}_2\hat{N'}\,-\mu \hat{H}_1\hat{H}_2-\epsilon_\alpha \hat{L}_\alpha \hat{H}_2\,
\ee
with 
\be \label{muep}
\epsilon_\alpha=-\lambda_\alpha  Us_\phi~~~,~~~\mu=-\kappa U c_\phi~.\ee

The model contains ten neutral fermion fields which mix with each other through the gauge and superpotential interactions. The corresponding mass Lagrangian in the basis $\psi_0 = (\nu_\alpha,N_\alpha, -i \lambda_1, -i \lambda_2, \tilde{h}_1^0,\tilde{h}_2^0 )^T$ is given by
\be \label{Lneut}
-{\cal L}_{\cal N} = \frac{1}{2}\, \psi_0^T\, {\cal M}_{\cal N}\, \psi_0\,+\,{\rm h.c.}\,,
\ee
where ${\cal M}_{\cal N}$ is a $10 \times 10$ mass matrix which can be parametrized as
\be \label{Mneut}
{\cal M}_{\cal N} = \left(\ba{ccc} 0 & m_D&m_{\nu \chi}  \\ 
                                                  m_{D}^T & M_N & m_{ N\chi} \\
                                                  m_{\nu\chi}^T & m_{ N\chi}^T & M_\chi \ea \right)\,.\ee
Here, $M_N$ is singlet neutrino mass matrix as given by Eq.  (\ref{MN}). $m_D$ is $3 \times 3$ Dirac neutrino mass matrix as can be read-off from Eq. (\ref{Wgenshifted}),
\be \label{mD}
m_D = v_2\, \left(\ba{ccc}
0&\lambda_e&0\\
0&\lambda_\mu&0\\
0&\lambda_\tau&0\\
\ea
\right)\,,
\ee
with $v_{1,2} \equiv \langle H_{1,2} \rangle$ and $\sqrt{v_1^2 + v_2^2} \equiv v = 174\,{\rm GeV}$. $m_{\nu \chi}$ and $m_{ N \chi}$ are $3 \times 4$  mass matrices. They can be determined from Eq. (\ref{Wgenshifted}) and the gauge interactions as
\be \label{mnuchi}
m_{\nu \chi} = \left(\ba{cccc} -\frac{1}{\sqrt{2}}g^\prime \omega^\prime_e & \frac{1}{\sqrt{2}}g \omega^\prime_e & 0 & -\epsilon_e \\
-\frac{1}{\sqrt{2}}g^\prime \omega^\prime_\mu & \frac{1}{\sqrt{2}}g \omega^\prime_\mu & 0 & -\epsilon_\mu \\
-\frac{1}{\sqrt{2}}g^\prime \omega^\prime_\tau & \frac{1}{\sqrt{2}}g \omega^\prime_\tau & 0 & -\epsilon_\tau \ea \right) \,,~~
m_{N\chi} = \left(\ba{cccc} 0 & 0 & 0 & 0 \\
0 & 0 & 0 & \lambda_\alpha \omega_\alpha^\prime  \\
0&0&\kappa v_2&\kappa v_1\\
\ea \right) \,,
\ee
where $\omega_\alpha' \equiv \vev{\tilde{\nu}_\alpha}$.
The explicit form of the neutralino mass matrix is
\be \label{Mchi}
M_{\chi} = \left( 
\ba{cccc}
 M_1 & 0 & -\frac{1}{\sqrt{2}} g^\prime v_1 &  \frac{1}{\sqrt{2}}g^\prime v_2 \\
 0 & M_2 & \frac{1}{\sqrt{2}} g v_1 & - \frac{1}{\sqrt{2}}g v_2 \\
-\frac{1}{\sqrt{2}} g^\prime v_1 & \frac{1}{\sqrt{2}} g v_1 & 0 & -\mu  \\
 \frac{1}{\sqrt{2}} g^\prime v_2 & -\frac{1}{\sqrt{2}} g v_2 & -\mu  & 0 \\ \ea \right)\,. \ee
The tree level light neutrino masses can be obtained from Eq. (\ref{Mneut}). For $m_{\nu \chi},m_D \ll M_\chi, M_N$, the neutrino mass matrix at the leading order is given by
\be \label{mnu}
m_\nu = -m_{3\times 7}\, {\cal M}_{7\times 7}^{-1}\,(m_{3\times 7})^T\,.\ee
Here, $m_{3\times 7}\equiv (m_D,m_{\nu\chi})$ is a $3\times 7$ analogue of the conventional Dirac matrix in familiar type I seesaw mechanism. ${\cal M}_{7\times 7}$ is a  mass matrix of heavier states corresponding to the lower $7\times 7$ block of Eq. (\ref{Mneut}).

Neglecting $\lambda_\alpha \omega_\alpha^\prime$ term\footnote{It is shown later in this section that the SM neutrino mass scale  requires $\lambda_\alpha \ll 1$ in this framework. Therefore, $\kappa v_{1,2} \gg \lambda_\alpha \omega^\prime_\alpha$ unless $\kappa$ is vanishingly small.} in $m_{N\chi}$, an explicit evaluation of Eq. (\ref{mnu}) gives
\beqa \label{mnu0_FI}
m_\nu & = & \tilde{m}_0 \left[\left(1-\frac{\kappa c_\phi U}{\mu}\left(2-\frac{\kappa c_\phi U}{\mu} \right) \right)\, s_\phi^2\, \tilde{\cal A} + \left(\frac{A_0 M_{N_1}}{v_2^2} -\frac{2\kappa^2 c_\phi^2 A_0 v_1}{\mu v_2}\right)\, \tilde{{\cal B}} \right. \nonumber \\
&-&  \left. \frac{\kappa A_0 c_\phi (v_1^2-v_2^2)}{v_2 \mu}\left(1-\frac{\kappa c_\phi U}{\mu}\right)\, s_\phi\,\tilde{{\cal H}} \right]\,,
\eeqa
where,
\be \label{m0tilde}
\tilde{m}_0 = \frac{D_4 v_2^2 U^2 \lambda^2}{\tilde{D}_7} = \frac{v_2^2}{M_{N_1} - \kappa^2 c_\phi^2 \zeta}\,, \ee
$D_4$ and $\tilde{D}_7$ are the determinants of $M_\chi$ and ${\cal M}_{7 \times 7}$, respectively. Explicit expression of $D_4$ is
\be \label{D4}
D_4={\rm Det}\,M_\chi = -\mu(M_1 M_2 \mu - (g^2 M_1 + g^{\prime 2} M_2) v_1 v_2)\,. \ee
$\tilde{D}_7$ can be obtained using the second equality in Eq. (\ref{m0tilde}) in terms of the other known parameters where
\be \label{MN1}
M_{N_1} \equiv 2 \epsilon \lambda U s_\phi c_\phi\,, \ee is the mass of the pGF and 
\be \label{zeta}
\zeta = \frac{v^4 (g^2 M_1 + g^{\prime 2} M_2) - 4 v_1 v_2 \mu M_1 M_2 }{2 D_4}\,.\ee
Further, 
\be \label{A0}
A_0 = - \frac{\mu (g^2 M_1 + g^{\prime 2} M_2)}{2 (M_1 M_2 \mu - v_1 v_2 (g^2 M_1 + g^{\prime 2} M_2))}\,.\ee
The matrices in Eq. (\ref{mnu0_FI}) are obtained as
\be \label{ABtilde}
\tilde{{\cal A}} =\left( \ba{ccc} \lambda_e^2 & \lambda_e \lambda_\mu & \lambda_e \lambda_\tau\\
\lambda_e \lambda_\mu & \lambda_\mu^2 &  \lambda_\mu \lambda_\tau\\
\lambda_e \lambda_\tau & \lambda_\mu \lambda_\tau & \lambda_\tau^2 \\ \ea \right)\,,~\tilde{{\cal B}} = \left( \ba{ccc} \omega_e^2 & \omega_e \omega_\mu & \omega_e \omega_\tau\\
\omega_e \omega_\mu & \omega_\mu^2 &  \omega_\mu \omega_\tau\\
\omega_e \omega_\tau & \omega_\mu \omega_\tau & \omega_\tau^2 \\ \ea \right)\,,
\ee
\be \label{Htilde}
\tilde{{\cal H}} = \left( \ba{ccc} 2 \lambda_e \omega_e & \lambda_e \omega_\mu + \lambda_\mu \omega_e & \lambda_e \omega_\tau + \lambda_\tau \omega_e\\
\lambda_e \omega_\mu + \lambda_\mu \omega_e & 2 \lambda_\mu \omega_\mu &  \lambda_\mu \omega_\tau + \lambda_\tau \omega_\mu \\
\lambda_e \omega_\tau + \lambda_\tau \omega_e &  \lambda_\mu \omega_\tau + \lambda_\tau \omega_\mu & 2 \lambda_\tau \omega_\tau \\ \ea \right)\,. \ee
Here, $\omega_\alpha$ denote the VEVs of sneutrino in the basis in which the bilinear term is rotated away from the superpotential \cite{Joshipura:2002fc}. Explicitly,
\be \label{omega}
\omega_\alpha =  \omega^\prime_\alpha - \frac{v_1}{\mu} \epsilon_\alpha\,.\ee

If one neglects mixing  $m_{N\chi}$ between the RH neutrino and neutralino induced by $\kappa$,  then the ${\cal M}_{7\times7}$ becomes block diagonal and $m_\nu$  becomes sum of two independent seesaw contributions corresponding to the first two terms of Eq. (\ref{mnu0_FI}) with $\kappa = 0$. The first term is the normal seesaw contribution while the second is the standard $R$-parity violating contribution arising due to neutrino-neutralino mixing. Both these contributions are of rank one and each can give mass to only one neutrino. As seen from Eq. (\ref{mnu0_FI}), the leading terms of the conventional seesaw contribution is solely governed by the mass of the pGF.

The $\kappa$-dependent terms in Eq. (\ref{mnu0_FI}) make significant contribution. This can be seen by eliminating $\kappa$ using the second in Eq. (\ref{muep}) which leads to
\beqa \label{mnu0_simp2}
m_\nu & = & \tilde{m}_0 \left[4 s_\phi^2\, \tilde{\cal A} + \frac{A_0 M_{N_1}}{v_2^2} \left(1 -\frac{2 \mu v_1 v_2}{U^2 M_{N_1}}\right)\, \tilde{{\cal B}} + \frac{2 A_0 (v_1^2-v_2^2)}{v_2 U}\, s_\phi\,\tilde{{\cal H}} \right]\,.\eeqa 
The snuetrino VEVs arise from the parameters $\epsilon_\alpha$ and one can parametrize them as 
\be \label{k}
\omega_\alpha \equiv k_\alpha \epsilon_\alpha = -\lambda_\alpha k_\alpha U s_\phi\,, \ee
where we have used Eq. (\ref{muep}). The elements of matrix $m_\nu$ in Eq. (\ref{mnu0_simp2}) can then be rewritten as
\be \label{mnu0_simp3}
\left(m_\nu\right)_{\alpha\beta}\ = \tilde{m}_0\, \lambda_\alpha \lambda_\beta\, s_\phi^2 \left[4 + \frac{A_0 M_{N_1} U^2}{v_2^2} \left(1 -\frac{2 \mu v_1 v_2}{U^2 M_{N_1}}\right)\,k_\alpha k_\beta - \frac{2 A_0 (v_1^2-v_2^2)}{v_2}\,(k_\alpha + k_\beta)\right]\,.\ee
Assuming the second and third terms are $\lesssim {\cal O}(1)$, one finds typically 
\be \label{lambda_order}
|\lambda_\alpha|^2 \sim {\cal O}\left(\frac{m_\nu M_{N_1}}{v_2^2}\right) \sim {\cal O}\left(10^{-13}\right) \times \left(\frac{M_{N_1}}{100\,{\rm GeV}}\right)\,. \ee 
Therefore, $|\lambda_\alpha| \ll 1$ and neglecting $\lambda_\alpha$ dependent terms in $m_{N \chi}$ is justified. The third term in Eq. (\ref{mnu0_simp3}) is smaller than the seesaw contribution for $k_\alpha \leq 1$. The second term however increases with $k_\alpha U$ and can compete with the seesaw contribution. Requiring that this term is at most of the order of the seesaw term implies
\be \label{kU_order}
U^2 |k_\alpha|^2 \lesssim {\cal O}\left(\frac{v_2^2}{|A_0| M_{N_1}}\right) \sim {\cal O}\left(10^6\right)\,{\rm GeV}^2 \times \left(\frac{2.8 \times 10^{-4}\,{\rm GeV}^{-1}}{|A_0|}\right) \times \left(\frac{100\,{\rm GeV}}{M_{N_1}}\right)\,. \ee 
The parameters $k_\alpha$ can be determined from the minimization of the scalar potential of the model, see Appendix \ref{AppA:kfactor} for details. Generically, $k_\alpha$ would be ${\cal O}(1)$. However, they can be very small if the soft SUSY breaking terms are correlated (see, Appendix \ref{AppA:kfactor}). For $k_\alpha \sim 1$, Eq. (\ref{kU_order}) implies a strong upper bound on the symmetry breaking scale which weakens with the decrease in the values of $k_\alpha$.

The magnitude of $k_\alpha$ depends on the trilinear soft terms and sneutrino soft masses along with the other parameters and in general it can be different for different $\alpha$. However, $k_\alpha$ can be identical in minimal SUGRA type theories because of flavour universal soft parameters at the scale of SUSY breaking mediation. A splitting among the values of $k_\alpha$ then get induced by renormalization group effects and it is expected to be small. Thus it is natural to assume flavour universal $k_\alpha$ in a large class of models. In such a situation, Eq. (\ref{mnu0_simp3}) can generate mass for only one neutrino and all the models in this category are ruled out in spite of the presence of various contributions to neutrino masses\footnote{This is not true if $\omega_\alpha$ and $\lambda_\alpha$
are regarded as independent. One can obtain correct masses and mixing  in this case, see for example, model in \cite{Lavignac:2020yld}. However, this amounts to assuming flavour dependent $k_\alpha$.}.

These models become viable only if sizeable flavour violations in $k_\alpha$ are introduced.  Even with the most general $k_\alpha$, Eq. (\ref{mnu0_simp3}) leads to a massless neutrino. We find that to get large enough solar angle and viable ratio of the solar to atmospheric squared mass differences, one requires $(k_2-k_3)/k_3 \simeq {\cal O}(1)$. Such a sizeable difference between $k_\alpha$ cannot come from the RG induced small corrections and one needs to assume large flavour violation in the soft terms at the SUSY breaking mediation scale. The need for such large flavour violation in case of the purely bilinear $R$-parity violation was noticed before \cite{Joshipura:2002fc,Romao:1999up,Hirsch:2000ef}. This is an unwelcome feature which can be cured if the flavour independent symmetry considered here is replaced by a flavour dependent leptonic symmetry. It is possible in this case to retain flavour universality of $k_\alpha$ and reproduce all the neutrino data as we discuss in the next section.

\section{$U(1)_{L_\mu-L_\tau}$ symmetry and neutrino masses}
\label{sec:lmultau}
The example we consider here is based on the spontaneously broken $U(1)_{L_\mu-L_\tau}$ symmetry. The three RH neutrinos $N_\alpha=(\hat{Y},\hat{N},\hat{N}^\prime)$ retain the same charges as before and we appropriately relabel them as a
$(\hat{N_e},\hat{N_\mu},\hat{N_\tau})$. The $U(1)$ charges for the three generations of the lepton doublet superfields, $\hat{L}_\alpha$ are  $(0,1,-1)$ while the $SU(2)$ singlet charged lepton superfields, $\hat{E}^c_\alpha$, have charges $(0,-1,1)$.  The MSSM Higgs superfields are neutral under $U(1)$. A renormalizable superpotential involving these superfields is written as 
\beqa \label{W}
W = \lambda_\alpha\, \hat{L}_\alpha \hat{H}_2 \hat{N}_\alpha\,+\,\lambda_\alpha^\prime\, \hat{L}_\alpha \hat{H}_1 \hat{E}^c_\alpha\,-\mu_0\hat{H}_1 \hat{H}_2\,+\kappa \hat{N}_e\, \hat{H}_1 \hat{H}_2\,
\eeqa
The above along with $W_N$ gives the complete superpotential characterizing the leptonic interactions in the model. Without loss of generality, the parameters $\lambda_\alpha$, $\lambda^\prime_\alpha$ and $\mu_0$ can be chosen real. The bilinear $R$-parity violation and the $\mu$ parameter generated through   the VEVs of scalar components of $\hat{N}_\alpha$ are given by 
\be \label{epmu2}
\epsilon_\alpha=-\lambda_\alpha\vev{\tilde{N}_\alpha}\,,~~\mu=\mu_0-\kappa \vev{\tilde{N}_e}\,\ee
The spontaneous breaking of $R$-parity also induces  nonzero VEVs for the sneutrino fields residing in $\hat{\nu}_\alpha$ at the minimum of the scalar potential.

The masses of all the ten neutral fermions are given by the matrix in Eq. (\ref{Mneut}). The matrices $M_N$, $M_{\nu\chi}$ and $M_\chi$ are the same as given before by Eqs. (\ref{MN},\ref{mnuchi},\ref{Mchi}), respectively. $m_D$ and $M_{N\chi}$ are different compared to the flavour independent cases and are given by
\be \label{mDf}
m_D = v_2\, {\rm Diag.}\left(\lambda_e, \lambda_\mu, \lambda_\tau \right)\,,~~~~
m_{N\chi} = \left(\ba{cccc} 
0 & 0 & \kappa v_2 & \kappa v_1+\lambda_e\omega_e'\\
0 & 0 & 0 & \lambda_\mu\omega_\mu'\\
0 & 0 & 0 & \lambda_\tau \omega_\tau'\\ \ea \right)\,. \ee
The tree level light neutrino mass matrix after the seesaw is still given by Eq. (\ref{mnu}). Neglecting $\lambda_\alpha \omega^\prime_\alpha$ terms in $M_{N \chi}$, an explicit evaluation of $m_\nu^{(0)}$ gives
\beqa \label{mnu0}
m_\nu^{(0)} & = & m_0 \left[{\cal A} + \frac{A_0 M_{N_1}}{v_2^2}\, {\cal B} +  \frac{2 \kappa \epsilon U}{\mu}\, {\cal C} +  \frac{\kappa A_0 \epsilon (v_1^2-v_2^2)}{v_2 \mu}\, {\cal D} \right. \nonumber \\
&+& \left. \frac{\kappa^2 \epsilon \zeta}{\lambda U}\,{\cal F} + \left(\frac{\kappa \epsilon U}{\mu}\right)^2\, {\cal G} - \frac{2\kappa^2 \epsilon^2 A_0 v_1}{\mu v_2}\,{\cal B} + \frac{\kappa^2 A_0 \epsilon^2 (v_1^2-v_2^2) U}{v_2 \mu^2}\,{\cal H} \right]\,,
\eeqa
where we have used relations Eqs. (\ref{omega},\ref{epmu2}). Further,
\be \label{m0}
m_0 =  \frac{D_4 v_2^2 U^2 \lambda^2}{D_7} = \frac{v_2^2}{M_{N_1} - \kappa^2 \epsilon^2 \zeta } \,,\ee
and $D_4$, $\zeta$ and $A_0$ are as defined earlier in Eqs. (\ref{D4},\ref{zeta},\ref{A0}) respectively. $D_7$ is the determinant of the new ${\cal M}_{7 \times 7}$ and it is different from $\tilde{D}_7$ defined earlier unless $\kappa = 0$. 
Also, 
\be \label{AB}
{\cal A} = \left( \ba{ccc} \epsilon^2 \lambda_e^2 & -\epsilon s_\phi \lambda_e \lambda_\mu & - \epsilon c_\phi \lambda_e \lambda_\tau \\
 -\epsilon s_\phi \lambda_e \lambda_\mu & s_\phi^2 \lambda_\mu^2 & - s_\phi c_\phi \lambda_\mu \lambda_\tau \\  
 - \epsilon c_\phi \lambda_e \lambda_\tau & - s_\phi c_\phi \lambda_\mu \lambda_\tau & c_\phi^2 \lambda_\tau^2 \ea \right)\,,~{\cal B} = \left( \ba{ccc} \omega_e^2 & \omega_e \omega_\mu & \omega_e \omega_\tau\\
\omega_e \omega_\mu & \omega_\mu^2 &  \omega_\mu \omega_\tau\\
\omega_e \omega_\tau & \omega_\mu \omega_\tau & \omega_\tau^2 \\ \ea \right)\,,
\ee
\be \label{C}
{\cal C} = \left( \ba{ccc} \epsilon^2 \lambda_e^2 & 0 & 0 \\
0 & - s_\phi^2 \lambda_\mu^2 & - s_\phi c_\phi \lambda_\mu \lambda_\tau \\  
0 & - s_\phi c_\phi \lambda_\mu \lambda_\tau & - c_\phi^2 \lambda_\tau^2 \ea \right)\,,
\ee
\be \label{D}
{\cal D} = \left( \ba{ccc} 2 \epsilon \lambda_e \omega_e & \epsilon \lambda_e \omega_\mu - s_\phi \lambda_\mu \omega_e  & \epsilon \lambda_e \omega_\tau - c_\phi \lambda_\tau \omega_e\\
\epsilon \lambda_e \omega_\mu - s_\phi \lambda_\mu \omega_e & -2 s_\phi \lambda_\mu \omega_\mu &  - s_\phi \lambda_\mu \omega_\tau - c_\phi \lambda_\tau \omega_\mu\\
\epsilon \lambda_e \omega_\tau - c_\phi \lambda_\tau \omega_e &  - s_\phi \lambda_\mu \omega_\tau - c_\phi \lambda_\tau \omega_\mu & -2 c_\phi \lambda_\tau \omega_\tau \\ \ea \right)\,,
\ee
\be \label{FG}
{\cal F} = \left( \ba{ccc} 0 & 0 & 0 \\
 0 & 0 & \lambda_\mu \lambda_\tau \\  
0 & \lambda_\mu \lambda_\tau & 0 \ea \right)\,,~~
{\cal G} = \left( \ba{ccc} \epsilon^2 \lambda_e^2 & \epsilon s_\phi \lambda_e \lambda_\mu & \epsilon c_\phi \lambda_e \lambda_\tau \\
 \epsilon s_\phi \lambda_e \lambda_\mu & s_\phi^2 \lambda_\mu^2 & s_\phi c_\phi \lambda_\mu \lambda_\tau \\  
 \epsilon c_\phi \lambda_e \lambda_\tau & s_\phi c_\phi \lambda_\mu \lambda_\tau & c_\phi^2 \lambda_\tau^2 \ea \right)\,,
\ee
\be \label{H}
{\cal H} = \left( \ba{ccc} 2 \epsilon \lambda_e \omega_e & \epsilon \lambda_e \omega_\mu + s_\phi \lambda_\mu \omega_e  & \epsilon \lambda_e \omega_\tau + c_\phi \lambda_\tau \omega_e\\
\epsilon \lambda_e \omega_\mu + s_\phi \lambda_\mu \omega_e & 2 s_\phi \lambda_\mu \omega_\mu &  s_\phi \lambda_\mu \omega_\tau + c_\phi \lambda_\tau \omega_\mu\\
\epsilon \lambda_e \omega_\tau + c_\phi \lambda_\tau \omega_e &   s_\phi \lambda_\mu \omega_\tau + c_\phi \lambda_\tau \omega_\mu & 2 c_\phi \lambda_\tau \omega_\tau \\ \ea \right)\,. \ee

In the limit $\kappa \to 0$, the first term in Eq. (\ref{mnu0}) quantifies the standard seesaw contribution: 
\be \label{mnuS}
m_{\nu}^S \equiv -m_D\, M_N^{-1}\, m_D^T = \frac{v_2^2}{M_{N_1}}\,{\cal A}\,.\ee
Contrary to the flavour independent cases, this contribution leads to non-zero masses for all three neutrinos. The leading mass 
\be \label{}
m_{\nu_3} \simeq \frac{v_2^2}{|M_{N_1}|}\left(c_\phi^2 \lambda_\tau^2 + s_\phi^2\lambda_\mu^2\right)\,, \ee
is controlled by the mass of the pGF. The lower $2\times 2$ block of the matrix ${\cal A}$ along with this mass can account for the atmospheric neutrino oscillations. The other two masses are degenerate when the terms of ${\cal O}(\epsilon^2)$ are neglected in ${\cal A}$. Degeneracy is lifted by the latter but by itself it is not sufficient to account for the solar scale and angle. The seesaw term in Eq. (\ref{mnu0}) however gets significant contribution from the $\kappa$ dependent terms ${\cal C}$, ${\cal F}$ and ${\cal G}$ when $\left|\kappa U \epsilon / \mu \right| \sim {\cal O}(1)$. These terms by themselves are sufficient to reproduce the neutrino spectrum as we will discuss. For $\kappa = 0$, the second term denotes the usual contribution to the neutrino masses from the sneutrino VEVs induced by $R$-parity violation:
\be \label{mnuR}
m_{\nu}^{\cancel{R}} \equiv - m_{\nu \chi}\, M_\chi^{-1}\, m_{\nu \chi}^T = A_0\,{\cal B}\,.\ee
This contribution also gets corrected by the $\kappa$ dependent terms as shown in Eq. (\ref{mnu0}).

As mentioned earlier, the parameters $\omega_\alpha$ denote the VEVs of sneutrino in the basis in which the bilinear term is shifted away \cite{Joshipura:2002fc}. They can be parametrized as $\omega_\alpha = k_\alpha \epsilon_\alpha$
and $k_\alpha$ can be determined from minimization of the scalar potential of the model, see Appendix \ref{AppA:kfactor} for details. The magnitude of $k_\alpha$ depends on the trilinear soft terms and sneutrino soft masses along with the other parameters and in general their value can be different for different $\alpha$. However, $k_\alpha$ are expected to be universal in minimal SUGRA type theories because of flavour universal soft parameters at the scale of SUSY breaking mediation. A splitting among the values of $k_\alpha$ then get induced by renormalization group effects. We assume that these effects are small and $k_\alpha$ are approximately universal at the electroweak scale as well. For $\omega_\alpha = k \epsilon_\alpha$, we find
\be \label{matrel}
{\cal B} = k^2 U^2\, {\cal G}\,,~~{\cal D} = -2 k U\, {\cal C}\,,~~{\cal H}=-2 k U\, {\cal G}\,.\ee
Substituting these in Eq. (\ref{mnu0}) and defining $r_{e,\mu} = \lambda_{e,\mu}/\lambda_\tau$, we obtain the neutrino mass matrix after combining the different contributions as
\be \label{mnu0_tot}
m^{(0)}_\nu = m_0  \lambda_\tau^2 \left(
\begin{array}{ccc}
 \epsilon ^2 r_e^2 (1 + \eta + \eta^\prime) & - \epsilon  r_e  r_\mu s_\phi (1 - \eta) & - \epsilon  r_e c_\phi (1 - \eta) \\
 -\epsilon  r_e  r_\mu s_\phi (1 - \eta) & r_\mu^2 s_\phi^2 (1 + \eta - \eta^\prime) & - r_{\mu } s_\phi c_\phi (1 - \eta + \eta^\prime - \zeta^\prime) \\
 - \epsilon  r_e c_\phi (1 - \eta)  & - r_{\mu } s_\phi c_\phi (1 - \eta + \eta^\prime - \zeta^\prime) & c_\phi^2  (1 + \eta - \eta^\prime) \\
\end{array}
\right),\ee
where,
\beqa \label{eta_etc}
\eta &=& \frac{A_0 k^2 U^2 M_{N_1}}{v_2^2} + \left( \frac{\kappa \epsilon U}{\mu}\right)^{2} \left(1 - \frac{2 A_0 k (k v_1 \mu + v_1^2-v_2^2)}{v_2}\right)\,,\nonumber \\
\eta^\prime &=& \frac{2 \kappa \epsilon U}{\mu} \left(1 - \frac{A_0 k (v_1^2-v_2^2)}{v_2}\right)\,,\nonumber \\
\zeta^\prime &=& \frac{2 \kappa^2 \epsilon^2 \zeta}{M_{N_1}}\,.\eeqa

If the induced sneutrino VEVs $\omega_\alpha$ are small then only the seesaw terms contribute to Eq. (\ref{mnu0_tot}). This contribution by itself is sufficient to describe neutrino masses and mixing parameters including the Dirac CP phase. However, it requires relatively low value of the $U(1)$ breaking scale $U$. This is quantified as follows. One obtains a relation $\eta=\frac{1}{4}\eta^{\prime 2}$ from Eq. (\ref{eta_etc}) in the purely seesaw limit. Eq. (\ref{matrel}) is then described in terms of five real parameters $m_0 \lambda_\tau^2$, $\eta^\prime$, $\phi$, $r_e \epsilon$, $r_\mu$ and a complex  $\zeta^\prime$ which control neutrino masses and mixing in this case. Explicitly,
\beqa \label{seesawnumbers}
r_e \epsilon &=& -0.649\,,~~r_\mu = -13.23\,,~~\eta^\prime = 0.714\,,~~\phi = -0.09\,,\nonumber \\
 \zeta^\prime &=& 0 .516 + 0.022\, i\,,~~m_0 \lambda_\tau^2 = 0.0367\, {\rm eV}\,,\eeqa
reproduce all the neutrino squared mass differences and mixing angles within less than 1$\sigma$ and the Dirac CP phase within 1.4$\sigma$ ranges given by the latest global fit results, NuFIT v5.0 \cite{Esteban:2020cvm}. The implied value of $\eta^\prime = \frac{2 \kappa \epsilon U}{
\mu} = \frac{2 \kappa \vev{\tilde{N}_e}}{\mu}$ determines the VEV of $\tilde{N}_e$ induced by the supergravity effect. The definitions of $\zeta^\prime$ and $M_{N_1}$ can be used to obtain a relation
\be \label{}
U^2 =\left| \frac{\kappa \eta^\prime \mu \zeta}{\lambda \zeta^\prime \sin2\phi}\right|\,.\ee
The values in Eq. (\ref{seesawnumbers}) then imply
\be \label{}
U = 432\,{\rm GeV}\, \times \sqrt{\left|\frac{\kappa}{\lambda}\right|}\, \times \left(\frac{\mu}{\rm TeV}\right)^{1/2}  \times\,\left(\frac{\zeta}{24.13\, {\rm GeV}}\right)^{1/2}\,.\ee
corresponding to a relatively low breaking of the $U(1)$ symmetry. The pGF mass is determined as
\be \label{}
|M_{N_1}|=\left|\frac{\lambda}{\kappa}\,\eta^\prime \mu\, s_\phi c_\phi \right| = 64\, {\rm GeV} \,\times \left|\frac{\lambda}{\kappa}\right| \times \left(\frac{\mu}{\rm TeV}\right)\,,\ee
while the heavier two neutrinos have masses $ 432 \sqrt{|\lambda \kappa|}\,{\rm GeV} \mp \frac{1}{2}|M_{N_1}|$. 

The $R$-parity violating contribution to the seesaw model can be significant with increasing $U$ and need to be included. Also, Eq. (\ref{mnu0_tot}) is a neutrino mas matrix as predicted by the model at the tree level. The neutrino masses also receive higher order corrections and for the masses of RH neutrinos of ${\cal O}(100)$ GeV, these corrections are non-negligible in general. We, therefore, include 1-loop correction to determine the viability of this model in reproducing the neutrino masses and mixing parameters and to derive its reliable predictions. There are several sources of 1-loop correction to the neutrino masses in this model. The dominant contribution arises from the self-energy correction mediated by the lightest Higgs $H$ and $Z$ bosons \cite{Grimus:2002nk,AristizabalSierra:2011mn}. These are evaluated in \cite{Grimus:2002nk} for multi Higgs doublets. Using this we obtain in our case, 
\be \label{dmnu}
\delta m_\nu^{(1)} =\frac{1}{32 \pi^2 v^2}\, m_D\,U_N^*\,{\rm Diag.}\left(g(M_{N _1}),g(M_{N_2}),g(M_{N_3})\right)\, U_N^\dagger\, m_D^T\,, \ee 
where $U_N$ is a unitary matrix which diagonalizes $M_N$ such that 
\be \label{}
M_N = U_N {\rm Diag.}(M_{N_1},M_{N_2},M_{N_3}) U_N^T\,, \ee
and
\be \label{g}
g(M_i) = M_i \left(\frac{1}{\sin^2\beta}\,\frac{\log \frac{M_i^2}{m_H^2}}{\frac{M_i^2}{m_H^2}-1}\, +\, 3\,\frac{\log \frac{M_i^2}{m_Z^2}}{\frac{M_i^2}{m_Z^2}-1} \right)\,.\ee
Similar corrections at 1-loop also arise from supersymmetric particles in the loop. We list them in Appendix \ref{AppB:loop}. Complete determination of the magnitude of these correction in general requires specification of SUSY breaking sector and depending on the later their contribution can be sizeable or sub-dominant. For a range of parameters of our interest, we find that these contributions remain small and $\delta m_\nu^{(1)}$ given in Eq. (\ref{dmnu}) captures the most relevant and dominant 1-loop correction.

The 1-loop corrected neutrino mass matrix is therefore given by
\be \label{mnu_tot}
m_\nu = m^{(0)}_\nu + \delta m_\nu^{(1)}\,. \ee
The charged lepton mass matrix is dominantly given by $3 \times 3$ matrix $m_l$ which is diagonal and real. Hence, the lepton mixing matrix can be determined from diagonalization of $m_\nu$ only. Note that $\lambda_\alpha$, $\epsilon$ and $U$ can be chosen real without loss of generality. $A_0$ and $k$ are also real if no CP violation is introduced in the soft SUSY breaking sector. The CP violation in the lepton sector, therefore, arises only from a parameter $\lambda$. Considering the present generic limits on the masses of neutralinos and charginos \cite{Zyla:2020zbs}, we take $M_1 \simeq M_2 \simeq \mu = 1$ TeV to evaluate the neutrino mass spectrum. We also fix $\tan\beta \equiv v_2/v_1 = 2$. This implies $\zeta = 24.13$ GeV from Eq. (\ref{zeta}) and $A_0 = -2.8 \times 10^{-4}\, {\rm GeV^{-1}}$ from Eq. (\ref{A0}). We fix $\kappa = 1$ and choose four sample values for $U$ and determine the remaining dimensionless eight real parameters (real $r_e$, $r_\mu$, $\epsilon$, $\phi$, $\lambda_\tau$, $k$ and complex $\lambda$) using $\chi^2$ minimization. The $\chi^2$ function includes two neutrino squared mass differences, three mixing angles and a Dirac CP phase. We use the values of these observables from the latest global fit results (NuFIT v5.0) \cite{Esteban:2020cvm}. We also assume normal ordering in neutrino masses.

The results for four example values of $U$ are displayed in Table \ref{tab:sol}. 
\begin{table}[t]
\begin{center}
\begin{tabular}{ccccc} 
\hline
\hline
 ~~~~~~Parameters ~~~~~~& ~~~~~~BP 1~~~~~~  & ~~~~~~BP 2~~~~~~ & ~~~~~~BP 3~~~~~~ & ~~~~~~BP 4~~~~~~ \\
\hline
$U$ [GeV] & $10^3$ & $10^4$  & $10^5$ &  $10^6$ \\
$\kappa$ & $1.0$ & $1.0$  & $1.0$ &  $1.0$ \\
$\lambda$ &  $0.5591 + 0.1449\,i$ & $0.6148 - 0.2204\,i$ & $0.7513 + 0.0327 \,i$ & $-0.8436 - 0.0283\,i$ \\
$\epsilon$ & 0.1648 & -0.0955 & 0.0276 & 0.0035\\
$\phi$ & 3.0965 & -3.011 & -2.8527 & -0.7434\\
$\lambda_\tau$ & $9.73 \times 10^{-8}$ & $2.321 \times 10^{-7}$ & $-5.895 \times 10^{-7}$  & $-1.2484 \times 10^{-6}$\\
$r_\mu$ & 27.2463 & -8.3797 & 4.614 & -1.5014\\
$r_e$ & -5.2322 & -140.889 & -16.6385 & -55.4134\\
$k$ & 0 & 0.0252 & -0.006 & 0.0005 \\
\hline
$\Delta m^2_{21}$ [$10^{-5}$ eV$^2$] & 7.42 & 7.35 & 7.42 & 7.42\\
$\Delta m^2_{31}$ [$10^{-3}$ eV$^2$] & 2.516 & 2.517  & 2.517  & 2.517\\
$\sin^2 \theta_{12}$ & 0.302 & 0.299 & 0.304 & 0.304\\
$\sin^2 \theta_{23}$ & 0.578 & 0.584 & 0.573 & 0.573 \\
$\sin^2 \theta_{13}$ & 0.0221 & 0.0221 & 0.0222 & 0.0222\\
$\sin \delta_{\rm CP}$ & -0.78 & -0.57 & -0.29 & -0.29\\
\hline
$\sum m_{\nu_i}$ [eV] & 0.146 & 0.086 & 0.09 & 0.092\\
$|m_{\beta \beta}|$  [eV] & 0.039 & 0.016 & 0.019 & 0.019\\
$M_{N_1}$ [GeV] & 8.34 & 159.7 & 1132  & 2912\\
$M_{N_2}$ [GeV] & 581.1 & 6479  & $7.466 \times 10^{4}$ & $8.426 \times 10^{5}$\\
$M_{N_3}$ [GeV] & 589.5 & 6639  & $7.579 \times 10^{4}$ & $8.455 \times 10^{5}$\\
\hline
\hline
\end{tabular}
\end{center}
\caption{The best fit values of model parameters, corresponding results for neutrino masses and mixing observables and relevant predictions for four benchmark solutions obtained for $\kappa = 1$ and different values of $U$.}
\label{tab:sol}
\end{table}
The solution BP 1 represent the pure seesaw case discussed earlier in Eq. (\ref{seesawnumbers}) but it now includes the radiatiove correction. This is shown here for specific values of various parameters for comparison. The benchmark points BP 3 and BP 4 reproduce the central values of neutrino squared mass differences and mixing parameters as obtained from the latest global fit results \cite{Esteban:2020cvm}. BP 1 and BP 2 reproduce all the observables within $1.3 \sigma$. It is seen that the 1-loop correction considered in Eq. (\ref{dmnu}) significantly modifies one of the elements of $m_\nu$ from its tree level value. This is shown explicitly in Appendix \ref{AppB:loop} where we also discuss the strengths of the other 1-loop corrections. We also repeat the same analysis but for $\kappa =0$ and the corresponding results are discussed in Appendix \ref{AppC:vanishing_kappa}.

In Table \ref{tab:sol}, we also list predictions for the sum of the three light neutrino masses, the effective mass parameter relevant for neutrinoless double beta decay (i.e. $m_{\beta \beta} = U_{ei}^2\, m_{\nu_i}$), and mass spectrum of the RH neutrinos. It can be seen that all the benchmark points predict relatively degenerate mass spectrum for the light neutrinos. The obtained value of $|m_{\beta \beta}|$ is smaller than the current upper limits from various recent experiments seearching for neutrinoless double beta decay as listed in \cite{Dolinski:2019nrj}. The strongest constraint on the neutrino mass spectrum of the model comes from the cosmological observations. The current limit from Planck collaboration is $\sum m_{\nu_i} < 0.12$ eV \cite{Aghanim:2018eyx} (see also \cite{Vagnozzi:2017ovm}) which seems to be in conflict with the prediction of BP 1. However, in deriving the constraint on the sum of the neutrino masses from cosmology, it is assumed that the neutrinos are stable or atleast their lifetime is greater than the age of the universe. If neutrinos decay into invisible radiation on the time scales shorter than the age of the universe, the above constraint can be  replaced by a much weaker limit \cite{Chacko:2019nej}, as weak as $\sum m_{\nu_i} < 0.9$ eV. In our model, neutrinos can decay into Majoron allowing such a possibility. We compute the neutrino lifetime in the next section and show that the solution BP 1 can evade the strongest limit\footnote{Alternatively, the stricter limit on $\sum m_{\nu_i}$ can also be relaxed in models based on dynamical dark energy with arbitrary equation of state \cite{Vagnozzi:2018jhn}.}.

A pair of the heavier neutrinos is quasi-degenerate while the mass of the pGF is predicted to be $ < {\cal O}(200)$ GeV for some of the benchmark points. The later is within the kinematic reach of the LHC. However, it's production and detection involves coupling with the SM particles which arise through active-sterile neutrino mixing in the standard scenario and hence the current bounds are sensitive to such mixing. We estimate this in section \ref{sec:direct} and obtain that the magnitude of the underlying mixing is $\lesssim 10^{-6}$. For such a small mixing, the current ATLAS \cite{Aad:2015xaa} and CMS \cite{Khachatryan:2015gha,Khachatryan:2016olu} analysis do not put any limit on the RH neutrino masses.

\section{Majoron and its couplings}
\label{sec:Majoron}
As a consequence of spontaneous  breaking of $U(1)_{L_\mu-L_\tau}$, there exists a massless Goldstone boson - Majoron - which was primarily given by the linear combination Eq.  (\ref{GB}). Since two of the three lepton doublets are also charged under $U(1)_{L_\mu-L_\tau}$, this structure gets modified to
\be \label{J}
J \simeq \cos \phi\, \tilde{N}_{I\tau} - \sin \phi\, \tilde{N}_{I\mu}+ \frac{\omega^\prime_\mu}{U}\, \tilde{\nu}_{I \mu} - \frac{\omega^\prime_\tau}{U}\, \tilde{\nu}_{I \tau}\,.
\ee
Majoron couples to the leptons at the tree level in this model while the couplings with the other SM particles can arise through loop(s). The leptonic Majoron couplings are strongly constrained from cosmology and astrophysics. 

The strongest constraints on the neutrino-Majoron couplings, denoted by $g_{i j}$ here, come from the observed  CMB spectrum which requires that  the neutrinos should be free streaming at the time of the photon decoupling. This restricts both the diagonal and the off-diagonal couplings \cite{Hannestad:2005ex,Raffelt:1994ry,Ayala:2014pea}. For the diagonal neutrino couplings with Majoron, the constraint reads $|g_{ii}| \lsim 10^{-7}$ while the off-diagonal couplings are required to be 
\be\label{gnuij}
|g_{i j}| \lsim 0.61\times10^{-11}\,\left(\frac{0.05\,{\rm eV}}{m_{\nu_i}}\right)^2\ee
for the hierarchical neutrino masses. 
The off-diagonal couplings $h_{\alpha\beta}$ of Majoron to the charged leptons are constrained from the non observation of the decay $e_\alpha\rightarrow e_\beta J\,.$ The branching ratio of $\alpha$ lepton decaying into Majoron and $\beta$ lepton is estimated as 
\be \label{}
{\rm BR}[\alpha \to \beta\, J ] = \frac{\Gamma[\alpha \to \beta\, J ]}{\Gamma[\alpha \to \beta\, \nu_\alpha\, \bar{\nu}_\beta]} \simeq \frac{12 \pi^2}{G_F^2 m_\alpha^4}\,|h_{\alpha \beta}|^2\,,
\ee
for $m_\alpha \gg m_\beta, m_J$.  The experimental limits ${\rm BR}[\mu \to e\, J] \le 2.6 \times 10^{-6}$ \cite{PhysRevD.34.1967} and ${\rm BR}[\tau \to e/\mu\, J] \le 10^{-3}$ \cite{Albrecht:1995ht} require
\be \label{hbound}
|h_{\mu e}| < 1.9 \times 10^{-11}\,,~~|h_{\tau e}|, |h_{\tau \mu}| < 1.1\times  10^{-7}\,.
\ee
The electron coupling to Majoran is contained from the energy loss from red giant and require \cite{Raffelt:1994ry,Ayala:2014pea}
\be \label{hee}
 |h_{ee}| < 2.57\times 10^{-13}\,.\ee  

We systematically evaluate leptonic couplings to Majoron below and show that the above bounds are satisfied  for the set of parameters which give the correct neutrino spectrum. In particular, the Majoron coupling to electron is absent in the model  in the seesaw limit and $g_{ee}$ much smaller than the value required above arise from higher order terms in the seesaw expansion.

\subsection{Couplings with neutrinos}
In the $U(1)_{L_\mu-L_\tau}$ based model, the dominant interaction of the light neutrinos with Majoron arises from primarily two sources: (a) through light-heavy neutrino mixing from $W_N$ in Eq.  (\ref{WN0}), and (b) through neutrino-neutralino mixing from $W$ in Eq. (\ref{W}) and gauge interactions. 

From the F-term of $W_N$ and substitution given in Eq.  (\ref{repl_N}), the relevant interaction terms are
\be \label{nuJa1}
{\cal L}^{(a)}_{\nu J} =- \frac{i \lambda}{\sqrt{2}}\, \left( N_\tau^T C^{-1} N_e\, \tilde{N}_{I \mu}  + N^{T}_\mu C^{-1} N_e\, \tilde{N}_{I \tau}\right) + {\rm h.c.}\,. \ee
The above gives rise to coupling between Majoron and the light neutrinos through heavy-light neutrino mixing. The later can be computed using diagonalization of the relevant blocks of neutral fermion mass matrix given in Eq.  (\ref{Mneut}) with $M_N$, $m_{\nu\chi}$, $M_\chi$ as written in Eqs. (\ref{MN},\ref{mnuchi},\ref{Mchi}) respectively and $m_D$, $m_{N \chi}$ given in Eq. (\ref{mDf}). The full matrix ${\cal M}_{\cal N}$ is block diagonalized at the leading order using a unitary matrix
\be \label{Uneut}
U_{\cal N} = \left( \ba{ccc} 1  & \rho_{D} & \rho_{\nu \chi}\\ 
								-\rho_D^\dagger & 1 & \rho_{N \chi}\\
								-\rho_{\nu \chi}^\dagger & -\rho_{N \chi}^\dagger & 1 \ea \right)\,+\, {\cal O}(\rho^2)\,.
\ee
such that $U_{\cal N}^T {\cal M}_{\cal N} U_{\cal N}$ is an approximate block diagonal matrix. Here,
\be \label{rhos}
\rho_{D} \simeq (m_D M_N^{-1})^*\,,~~
\rho_{\nu \chi} \simeq (m_{\nu \chi} M_\chi^{-1})^*\,,~~
\rho_{\chi N} \simeq (m_{N \chi} M_\chi^{-1})^*\,.\ee
The neutral fermions in the block-diagonal basis are then obtained from the original basis as $U_{\cal N}^\dagger \psi^0$. Using this and denoting the light neutrinos in block diagonal basis as $\nu_\alpha^\prime$, we find
\beqa \label{Nnu_mixing}
N_e &=& -(\rho_D^\dagger)_{1 \alpha} \, \nu_{L \alpha}^\prime = (-1)^{1-\delta_{\alpha 1}}\,B_0\, \epsilon\, \epsilon_\alpha \, \nu_{L \alpha}^\prime \,, \nonumber \\
N_\mu &=& -(\rho_D^\dagger)_{2 \alpha}\, \nu_{L \alpha}^\prime = (-1)^{1-\delta_{\alpha 2}}\, B_0\, \sin\phi\, \epsilon_\alpha \, \nu_{L \alpha}^\prime \,, \nonumber \\
N_\tau &=& -(\rho_D^\dagger)_{3 \alpha}\, \nu_{L \alpha}^\prime = (-1)^{1-\delta_{\alpha 3}}\, B_0\, \cos\phi\, \epsilon_\alpha \, \nu_{L \alpha}^\prime \,, \eeqa
where 
\be \label{B0}
B_0 = \frac{v_2}{U^2 \lambda \epsilon \sin2\phi} = \frac{v_2}{M_{N_1} U}\,.\ee

Substituting Eq. (\ref{Nnu_mixing}) in Eq. (\ref{nuJa1}) and use of Eq. (\ref{J}) determine the couplings of heavy and light neutrinos with Majoron. As we are interested in the latter, the relevant part is obtained as
\be \label{nuJa2}
{\cal L}^{(a)}_{\nu J} = -i \frac{B_0}{\sqrt{2}U}\, \frac{v_2}{U}\, (-1)^{(1-\delta_{\beta 1})}\, \epsilon_\beta \left(\epsilon_\mu\, \nu_{L \mu}^{\prime T} - \epsilon_\tau\, \nu_{L \tau}^{\prime T} \right) C^{-1} \nu_{L \beta}^\prime\, J\, + {\rm h.c.}\,.\ee
In the physical basis of the light neutrinos, $\nu_{L \alpha}^\prime = U_{\alpha i}\, \nu_{L i}$ where $U_{\alpha i}$ are the elements of the PMNS matrix, the above term is written as
\be \label{nuJa3}
{\cal L}^{(a)}_{\nu J} = -i g_{ij}^{(a)}\, \nu^T_{L i} C^{-1} \nu_{L j}\, J\, + {\rm h.c.}\,,\ee
with
\be \label{ga}
g^{(a)}_{ij} = \frac{B_0}{\sqrt{2}U}\, \frac{v_2}{U}\, \sum_{\beta}\, (-1)^{(1-\delta_{\beta 1})}\, \left(\epsilon_\mu\,U_{\mu i} -  \epsilon_\tau\,U_{\tau i} \right)\,  \epsilon_\beta\,U_{\beta j}\,.\ee

We now evaluate the neutrino-Majoron coupling which arise from the neutrino-neutralino mixing. The relevant Lagrangian can be read from Eq. (\ref{W}) and gauge interaction as
\beqa \label{nuJb1}
{\cal L}^{(b)}_{\nu J} &=& - \frac{i}{\sqrt{2}} \left( \lambda_\mu\, \nu_{L \mu}^T \tilde{N}_{I \mu}+ \lambda_\tau\, \nu_{L \tau}^T \tilde{N}_{I \mu} \right) C^{-1} \tilde{h}_2^0 + \frac{i}{2} (g^\prime \tilde{B}^T - g \tilde{W}^{0T}) C^{-1} \nu_{L \beta}\, \tilde{\nu}_{I \beta}+{\rm h.c.}\,, \eeqa
where we define $-i\lambda_1 \equiv \tilde{B}$ and $-i\lambda_2 \equiv \tilde{W}^0$. The neutral fermions $\tilde{h}_2^0$, $\tilde{B}$ and $\tilde{W}^0$ in the above contain the light neutrinos when ${\cal M}_{\cal N}$ is block diagonalized. From Eqs. (\ref{Uneut},\ref{rhos}), we obtain
\beqa \label{chinu_mixing}
\tilde{h}_2^0 &=& -(\rho_{\nu \chi}^\dagger)_{4 \alpha} \, \nu_{L \alpha}^\prime = \frac{v_1}{\mu} A_0\, \omega_\alpha \, U_{\alpha i}\, \nu_{L i}\,, \nonumber \\
\tilde{B} &=& -(\rho_{\nu \chi}^\dagger)_{1 \alpha}\,\nu_{L \alpha}^\prime = - \frac{\sqrt{2} g^\prime M_2}{g^2 M_1 + g^{\prime 2} M_2}\,A_0\, \omega_\alpha\, U_{\alpha i}\, \nu_{L i} \,, \nonumber \\
\tilde{W}^0 &=& -(\rho_{\nu \chi}^\dagger)_{2 \alpha}\, \nu_{L \alpha}^\prime = \frac{\sqrt{2} g M_1}{g^2 M_1 + g^{\prime 2} M_2}\,A_0\, \omega_\alpha\, U_{\alpha i}\, \nu_{L i} \,, \eeqa
where $A_0$ and $\omega_\alpha$ are defined in Eqs. (\ref{A0},\ref{omega}). Substituting the above relations in Eq. (\ref{nuJb1}) and use of Eq. (\ref{J}) lead to the following interaction between the SM neutrinos and Majoron:
\be \label{nuJb2}
{\cal L}^{(b)}_{\nu J} = -i g_{ij}^{(b)}\, \nu^T_{L i} C^{-1} \nu_{L j}\, J\, + {\rm h.c.}\,,\ee
with
\be \label{gb}
g^{(b)}_{ij} = \frac{A_0}{\sqrt{2}U}\, \sum_{\beta}\, \left(\omega_\mu\,U_{\mu i} -  \omega_\tau\,U_{\tau i} \right)\,  \omega_\beta\, U_{\beta j}\,.\ee

Combining the two contributions, the light neutrino-Majoron coupling is parametrized as
\be \label{nuJ}
-{\cal L}_{\nu J} = i g_{ij}\, \nu^T_{L i} C^{-1} \nu_{L j}\, J\, + {\rm h.c.}\,,\ee
with
\be \label{gnuJ}
g_{ij} = g_{ij}^{(a)} + g_{ij}^{(b)} = \frac{B_0 v_2}{\sqrt{2}\, U^2}\, \sum_\beta \left((-1)^{(1-\delta_{\beta 1})} + \eta_0 \right) \left(\epsilon_\mu\,U_{\mu i} -  \epsilon_\tau\,U_{\tau i} \right)\,  \epsilon_\beta\, U_{\beta j} \,,\ee
where we have used $\omega_\alpha = k \epsilon_\alpha$ and 
\be \label{eta0}
\eta_0 = \frac{A_0 k^2 U^2 M_{N_1}}{v_2^2}\,.\ee
Note that for $\kappa \to 0$, $\eta_0$ denotes the relative strength of $R$-parity violating contribution compared to the standard seesaw contribution to the neutrino masses. For $|\eta_0| < 1$, the neutrino-Majoron coupling is dominantly determined by the light-heavy neutrino mixing while for  $|\eta_0| > 1$ it mainly arises from the neutrino-neutralino mixing.

The interaction in Eq.  (\ref{nuJ}) leads to neutrino decay into Majoron whenever it is kinematically allowed. The decay width can be estimated as
\be \label{Gamma}
\Gamma[\nu_j \to \nu_i\, J] = \frac{\lambda^{1/2}[m_{\nu j}^2,m_{\nu i}^2,m_J^2]}{16 \pi\, m_{\nu j}^3}\, |g_{ji}|^2\, \left(m_{\nu j}^2 + m_{\nu i}^2 - m_J^2 - 2 m_{\nu i} m_J \right)\,,
\ee 
where $\lambda[x,y,z] = x^2 + y^2 + z^2 - 2xy -2 yz- 2zx$. Assuming massless Majoron, we compute the total decay width of $\nu_3$ as $\Gamma_\nu \equiv \Gamma[\nu_3 \to \nu_2\, J] + \Gamma[\nu_3 \to \nu_1\, J] $ and the lifetime $\tau_\nu =1/\Gamma_\nu$ for the benchmark solutions listed in Table \ref{tab:sol}. The results are displayed in Table \ref{tab:nudecay} where we also list the obtained value of $|\eta_0|$ and the relevant off-diagonal and the largest diagonal couplings $|g_{ij}|$. 
\begin{table}[t]
\begin{center}
\begin{tabular}{ccccc} 
\hline
\hline
~~~Parameters ~~~& ~~~~~~BP 1~~~~~~  & ~~~~~~BP 2~~~~~~ & ~~~~~~BP 3~~~~~~ & ~~~~~~BP 4~~~~~~ \\
\hline
$|\eta_0|$ & 0 & 0.118  & 4.79 &  7.67 \\
${\rm Max.}\{|g_{ii}|\}$ & $1.2 \times 10^{-14}$ & $5.8 \times 10^{-15}$ & $1.8 \times 10^{-16}$ & $2.8 \times 10^{-17}$\\
$|g_{32}|$ & $3.3 \times 10^{-14}$ & $8.9 \times 10^{-16}$ & $4.5 \times 10^{-17}$ & $6.2 \times 10^{-18}$ \\
$|g_{31}|$ & $4.0 \times 10^{-14}$ & $1.4 \times 10^{-15}$ & $4.1 \times 10^{-17}$ & $4.9 \times 10^{-18}$ \\
$\tau_\nu$ [sec.] & $2.3 \times 10^{14}$ & $2.4 \times 10^{17}$ & $1.7 \times 10^{20}$ & $1.0 \times 10^{22}$\\
\hline
\hline
\end{tabular}
\end{center}
\caption{Predictions for the largest diagonal coupling of neutrinos with Majoron, off-diagonal couplings relevant for neutrino decays and the corresponding neutrino lifetime for the benchmark solutions.}
\label{tab:nudecay}
\end{table}
The neutrino lifetime turns out to be smaller than the age of the universe for solution BP 1 and hence the stringent limit on the sum of neutrino masses does not apply on this solutions as discussed earlier. The derived values of the neutrino-Majoron couplings for all the benchmark solutions obey the limits Eq. (\ref{gnuij}).

\subsection{Couplings with charged leptons}                                                                                                                                                                                                          
The chargino spectrum of the model consists of three pairs of the charged lepton fields and pairs of charged wino and Higgsino. In the basis $\psi_+ = \left(e^c_\alpha, - i \lambda^+, h_2^+\right)^T$ and  $\psi_- = \left(e_\alpha, -i \lambda^-, h_1^-\right)^T$, the charged fermion masses are given by 
\be \label{Lcharg}
-{\cal L}_{\cal C} = \psi_-^T\, {\cal M}_{\cal C}\, \psi_+\,+\, {\rm h.c.}\,,
\ee
with
\be \label{Mcharg}
{\cal M}_{\cal C} = \left(\ba{cc}  m_l  & m_{lC} \\ 
                                                 m_{lC}^\prime & M_C \ea \right)\,,\ee
and
\be \label{mlMC}
m_l = v_1\, {\rm Diag.}\left(\lambda_e^\prime, \lambda_\mu^\prime, \lambda_\tau^\prime \right)\,,~~M_C = \left(\ba{cc} M_2 & g v_2 \\ g v_1 & \mu \ea \right)\,,
\ee
\be \label{mlc}
m_{lC} = \left(\ba{cc} g \omega^\prime_e & \epsilon_e \\ g \omega^\prime_\mu & \epsilon_\mu \\ g \omega^\prime_\tau & \epsilon_\tau \ea \right)\,,~~m_{lC}^\prime = \left(\ba{ccc} 0 & 0 & 0 \\ -\omega^\prime_e \lambda^\prime_e & -\omega^\prime_\mu \lambda^\prime_\mu & -\omega^\prime_\tau \lambda^\prime_\tau \ea \right)\,.
\ee

From $W$ in Eq. (\ref{W}) and gauge interactions, we get the following interaction term involving the charged fermions and Majoron
\beqa \label{Llj}
{\cal L}_{lJ} &=& \frac{i}{\sqrt{2}}\, \left(\lambda_\mu\, \mu \tilde{N}_{I \mu} + \lambda_\tau\, \tau \tilde{N}_{I \tau} \right) \tilde{h}_2^+ + \frac{i}{\sqrt{2}} \left(\lambda^\prime_\mu\,\tilde{\nu}_{I \mu} \mu^{c} + \lambda^\prime_\tau\, \tilde{\nu}_{I \tau} \tau^{c} \right) \tilde{h}_1^-\,  \nonumber \\
& - & \frac{i}{\sqrt{2}}\, g \left(\tilde{\nu}_{I \mu}\, \mu +\tilde{\nu}_{I \tau}\,  \tau \right) (-i \lambda^+)\,+ \, {\rm h.c.}\,.
\eeqa
The chargino-charged lepton mixing can be evaluated by block diagonalizing the matrix ${\cal M}_{\cal C}$ given in Eq. (\ref{Mcharg}). For $m_l < m_{lC} < M_C$, this is done using the following leading order unitary matrices
\be \label{}
U_+ = \left( \ba{cc} 1 & (M_C^{-1}m_{lC}^\prime)^\dagger \\ -M_C^{-1} m_{lC}^\prime & 1 \ea \right)\,,~~U_- = \left( \ba{cc} 1 & (m_{lC} M_C^{-1})^* \\ -(m_{lC} M_C^{-1})^T & 1 \ea \right)\,.
\ee
such that 
\be \label{BD_cl}
U_-^T\, {\cal M}_{\cal C}\, U_+ \approx \left(\ba{cc} m_l - m_{lC}\,M_C^{-1}\,m_{lC}^\prime & 0 \\ 0 & M_C \ea \right)\,. \ee
The fields $\tilde{h}_2^+$, $\lambda^+$ and $\tilde{h}_1^-$ can then be written in the block diagonal basis of the charged leptons, $e^\prime_\alpha$ and $e^{c \prime}_\alpha$, as
\beqa \label{mix}
h_2^+ & = & -\left(M_C^{-1} m_{lC}^\prime\right)_{5 \alpha}\, e^{c \prime}_\alpha = \frac{M_2}{M_2 \mu - g^2 v_1 v_2}\,\omega^\prime_\alpha \lambda^\prime_\alpha\, e^{c \prime}_\alpha \,, \nonumber \\
-i \lambda^+ & = & -\left(M_C^{-1} m_{lC}^\prime\right)_{4 \alpha}\, e^{c \prime}_\alpha = -\frac{g v_2}{M_2 \mu - g^2 v_1 v_2}\,\omega^\prime_\alpha \lambda^\prime_\alpha\, e^{c \prime}_\alpha \,\,, \nonumber \\
h_1^- & = & -\left((M_C^{-1})^T m_{lC}^T\right)_{5 \alpha}\, e^{\prime}_\alpha = -\frac{M_2 \epsilon_\alpha - g^2 v_2 \omega_\alpha^\prime}{M_2 \mu - g^2 v_1 v_2}\,e^{\prime}_\alpha \, \,. \eeqa

Substitution of Eqs. (\ref{mix},\ref{J}) in Eq. (\ref{Llj}) leads to
\be \label{Llj2}
-{\cal L}_{lJ} = \frac{i}{\sqrt{2}} \frac{1}{U v_1}\, \left(\omega^\prime_\alpha m_\alpha \xi_\mu\, \mu^\prime\, e^{c \prime}_\alpha + \omega_\mu m_\mu \xi_\alpha\, e^\prime_\alpha\, \mu^{c \prime} - (\mu \to \tau) \right)\, J\,+\,{\rm h.c.}\,,\ee
where  $m_\alpha \simeq v_1 \lambda^\prime_\alpha$ are the charged lepton masses, and
\be \label{}
\xi_\alpha = \frac{M_2 \epsilon_\alpha - g^2 v_2 \omega^\prime_\alpha}{M_2 \mu - g^2 v_1 v_2}\,.\ee
As  can be seen from Eq. (\ref{BD_cl}) the charged lepton mass matrix, $M_l = m_l - m_{lC}\,M_C^{-1}m_{lC}'$, in the block diagonal basis is approximately diagonal since $m_l$ is diagonal.  The primed fields and corresponding physical states are related with each other by a unitary matrix very close to the identity matrix. Hence, the Majoron dominantly couples to the second and third generation charged leptons and its coupling to a pair of electrons arises only through $e-\alpha$ mixing $U_{e \alpha} \approx {\cal O} \left(\frac{\epsilon_e \lambda^\prime_\alpha \omega^\prime_\alpha}{m_e \mu}\right)$ induced by the seesaw correction  $m_{lC} M_C^{-1} m^\prime_{lC}$.

We estimate the magnitude of the coupling appearing in Eq. (\ref{Llj2}) for the benchmark solutions listed in Table \ref{tab:sol}. $\omega_\alpha^\prime$ can be expressed in terms of the known parameters using Eqs. (\ref{omega},\ref{k}). For $\mu,M_2 \gg v$, one finds
\be \label{}
h_{\alpha \beta} \equiv  \frac{\omega^\prime_\alpha m_\alpha \xi_\beta}{\sqrt{2} U v_1} \simeq \left(k + \frac{v_1}{\mu}\right) \frac{m_\alpha \epsilon_\alpha \epsilon_\beta}{\sqrt{2}\, v_1 \mu U}\,.\ee
The magnitudes of the largest value of $h_{\alpha \beta}$ obtained for different benchmark points are given in Table \ref{tab:clmajoron}. These are consistent with the experimental bounds given in Eq.(\ref{hbound}).
\begin{table}[t]
\begin{center}
\begin{tabular}{ccccc} 
\hline
\hline
~~~Parameters ~~~& ~~~~~~BP 1~~~~~~  & ~~~~~~BP 2~~~~~~ & ~~~~~~BP 3~~~~~~ & ~~~~~~BP 4~~~~~~ \\
\hline
${\rm Max.}\{|h_{\alpha \beta}|\}$ &  $10^{-17}$ & $10^{-15}$ & $10^{-14}$ & $10^{-13}$ \\
\hline
\hline
\end{tabular}
\end{center}
\caption{Maximum value of charged leptons' coupling with Majoron obtained for the different  benchmark solutions.}
\label{tab:clmajoron}
\end{table}
The strength of couplings is larger for solutions with the higher $U(1)$ breaking scale in contrast to neutrino-Majoron coupling. As mentioned earlier, the coupling of Majoron with electrons will be further suppressed and will be much smaller than $10^{-13}$ for all the solutions. This is in agreement with the astrophysical bounds from red giant \cite{Raffelt:1994ry,Ayala:2014pea} and SN 1987A \cite{Keil:1996ju}.

\section{Direct searches of pseudo-Goldstone fermion}
\label{sec:direct}
In the standard seesaw models, the direct searches of heavy neutrinos typically depend on  production mechanisms which involve light-heavy neutrino mixing. Therefore, the production cross section is suppressed for small mixings between the active and sterile neutrinos \cite{Pilaftsis:1991ug}. This picture changes significantly in the present framework as the lightest RH neutrino, which is a pseudo-Goldstone fermion, can be produced from the decays of neutralinos \cite{Lavignac:2020yld}. The production of Higgsino-like neutralino state is governed by electroweak interactions and it is not suppressed in general. The RH neutrinos mix with neutralinos through mass mixing term $m_{\chi N}$ in Eq. (\ref{Mneut}). If this mixing is large enough then $N_1$ is produced in the decays of Higgsinos. For $M_{N_1} \ll m_{\chi^0}$, the produced heavy neutrino is boosted and its subsequent decays into $W$ and charged leptons or $Z$ and light neutrinos give rise to displaced vertex signature. Such a signature has a better visibility due to less SM background and hence it improves significantly the prospects of detecting the pGF. Note that the decays of $N_1$ in the visible particles are still governed by the light-heavy neutrino mixing. The details of the displaced vertex signature of pGF of this kind are recently discussed in \cite{Lavignac:2020yld}.

The parameter of primary interest here is the mixing between RH neutrinos and neutralino. This can be estimated following the same procedure outlined in section IV-A. One finds from Eq. (\ref{Uneut}), $N_\alpha = (\rho_{N \chi})_{\alpha m}\, \psi_{0,m}$ with $m=7,..,10$. This leads to the following leading order mixing between the RH neutrinos and neutralinos in the block-diagonal basis:
\beqa \label{Nchi_mixing}
N_e & \simeq & \frac{2 \kappa A_0 v M_1 M_2}{\mu (g^2 M_1 + g^{\prime 2} M_2)}\,\left(\cos\beta\, \tilde{h}^0_1 + \sin\beta\, \tilde{h}^0_2 + \frac{g^\prime v}{\sqrt{2} M_1}\cos 2\beta\, \tilde{B} -\frac{g v}{\sqrt{2} M_2}\cos 2\beta\, \tilde{W^0}  \right). \eeqa
We have neglected $\lambda_\alpha$ dependent terms in $m_{N\chi}$ while deriving the above. $N_{\mu,\tau}$ do not contain any neutralino state at the leading order. Substituting the above in Eq. (\ref{pGF}), one can determine the mixing between the pGF $N_1$ and neutralinos. As it can be seen, $N_1$ has dominant mixing with the Higgsino states and they are obtained as 
\beqa \label{VNchi}
V_{{N_1} \tilde{h}^0_1} & = & -\left(\frac{2 A_0 v M_1 M_2}{\mu (g^2 M_1 + g^{\prime 2} M_2)}\right)\, \kappa \epsilon \cos2\phi\, \cos\beta\, \simeq \frac{v}{\mu}\kappa \epsilon \cos2\phi\, \cos\beta\,, \nonumber \\
V_{{N_1} \tilde{h}^0_2} &\simeq & \tan\beta\, V_{{N_1} \tilde{h}^0_1}\,,
\eeqa
for $v \ll M_{1,2}$. This can be compared with the $N_1$-$\nu_i$ mixing which can be read-off from similar expressions, Eq. (\ref{Nnu_mixing}), as
\be \label{Vnu}
V_{{N_1} {\nu_i}} = V_{{N_1} {\nu_\alpha}}\, U_{\alpha i}\,, \ee
where
\beqa \label{VNnu1}
V_{{N_1} {\nu_e}} &\simeq & - B_0\, \epsilon_e \cos2\phi\, (1+\epsilon^2)\,, \nonumber \\
V_{{N_1} {\nu_\mu}} &\simeq & - B_0\, \epsilon_\mu\, (1 - \epsilon^2 \cos2\phi)\,, \nonumber \\
V_{{N_1} {\nu_\tau}} &\simeq & B_0\, \epsilon_\tau\, (1 + \epsilon^2 \cos2\phi)\,.
\eeqa

For $\kappa \sim {\cal O}(1)$, one finds $| V_{{N_1}  \tilde{h}^0_{1,2}}| \gg | V_{{N_1}{\nu_i}}|$ as the later is suppressed by Yukawa couplings $\lambda_\alpha$. The estimated values of the above mixing for the different benchmark solutions are listed in Table \ref{tab:mixing}. 
\begin{table}[t]
\begin{center}
\begin{tabular}{ccccc} 
\hline
\hline
~~~Parameters ~~~& ~~~~~~BP 1~~~~~~  & ~~~~~~BP 2~~~~~~ & ~~~~~~BP 3~~~~~~ & ~~~~~~BP 4~~~~~~ \\
\hline
$\left|V_{{N_1} \tilde{h}^0_1}\right|$ & 0.013 & 0.0072 & 0.0007  & 0.0011 \\
$\left|V_{{N_1} {\nu_1}}\right|$ & $1.6 \times 10^{-6}$ &  $2.6 \times 10^{-6}$ & $2.7 \times 10^{-8}$  & $6.4 \times 10^{-8}$ \\
$\left|V_{{N_1} {\nu_2}}\right|$ & $1.0 \times 10^{-6}$ & $1.3 \times 10^{-6}$ & $3.2 \times 10^{-7}$  & $3.9 \times 10^{-7}$ \\
$\left|V_{{N_1} {\nu_3}}\right|$ & $2.5 \times10^{-6}$ & $4.7 \times 10^{-7}$ & $1.1 \times 10^{-7}$  & $2.6 \times 10^{-7}$ \\
\hline
\hline
\end{tabular}
\end{center}
\caption{Estimated magnitudes of the pseudo-Goldstone fermion's mixing with Higgsino and light neutrinos for the benchmark solutions.}
\label{tab:mixing}
\end{table}
For the first two solutions, the obtained values of mixings are such that the pGF with mass less than $200$ GeV can be detected through displaced vertex signature at the LHC \cite{Lavignac:2020yld}.

\section{Summary and Discussions}
\label{sec:summary}
Possibility of a RH neutrino being a Goldstone fermion of a spontaneously broken global $U(1)$ symmetry in the supersymmetric theories offers an interesting opportunity to confirm or otherwise, the mechanism of neutrino mass generation in direct search experiments. The singlet fermion is naturally light with mass of the order of gravitino mass scale in minimal supergravity like setup. Moreover, it can couple to some of the supersymmetric particles with large enough couplings such that the former can be produced at colliders through decays of the later. Investigating this general framework in detail, we find that the observed neutrino mass spectrum severely restricts the nature of $U(1)$ symmetry.

The light neutrino masses arise in the framework through the standard type I seesaw mechanism as well as through the $R$-parity violating RH snuetrino VEVs. Because of their common origin, these contributions are highly correlated. It is shown that if all three families of the SM leptons have the same $U(1)$ charge and the soft SUSY breaking terms are flavour universal then two of the three light neutrinos remain massless despite the presence of more than one sources of neutrino mass generation. One requires large flavour violations in the soft sector to induce the solar neutrino mass scale with one neutrino still remaining massless. We show that this picture changes significantly when the underlying $U(1)$ is flavour dependent symmetry. Considering a particular example of $L_\mu - L_\tau$ symmetry, it is shown that viable neutrino masses and mixing can be obtained without introducing any flavour violation in the soft sector.

The framework also contains a Majoron - the Goldstone boson of $U(1)_{L_\mu-L_\tau}$. We derive its couplings with the SM leptons and show that it couples weakly enough to evade the present constraints from cosmology, the stellar cooling as well as the constraints from the flavour violating decays of the charged leptons. Given this, and the fact that the viable neutrino mass spectrum can be achieved without any explicit breaking of $U(1)_{L_\mu-L_\tau}$, the Majoron can remain massless in the present framework. Light neutrinos can also decay into Majoron in this case and we find that the neutrino lifetime can be of the order of the age of universe or smaller if the scale of $U(1)$ breaking is $\lesssim\, 10^4$ GeV.

The nature of neutrino mass generation mechanism in this framework appears similar to the one in $\mu\nu$SSM model \cite{Lopez-Fogliani:2005vcg,Escudero:2008jg,Ghosh:2008yh,Fidalgo:2009dm}. In the later, the bilinear terms including the $\mu$ term are assumed to be absent at the begining and they arise from the trilinear terms when RH neutrinos take VEVs. Typically, the most general trilinear terms are considered which explicitly violate the $U(1)$ lepton number. As a result, both the Dirac and RH neutrino mass matrices take general form in $\mu\nu$SSM. In contrast to this, the present framework uses a flavour dependent combination of lepton number as a global symmetry which leads to highly restricted forms of $m_D$ and $M_N$ and still gives viable explanation of lepton spectrum. More importantly, the structure of the RH neutrino masses is determined here from the combined requirement of the unbroken global supersymmetry and the spontaneous violation of the $U(1)$ symmetry. The $U(1)$ symmetry protects the masses of a boson and its superpartner which is identified as the lightest RH neutrino. The later receives mass from the local SUSY breaking effects. A weak scale RH neutrino as pGF, therefore, naturally emerge in this setup. Thus, the formalism presented here is more constrained than the $\mu\nu$SSM. If $\mu_0 = 0$ set by some mechanism, this shift in VEVs also generates a TeV scale $\mu$ parameter in the present framework in a simlar way discussed recently in \cite{Dvali:2020uvd}. In this case, one typically finds $\mu \sim M_{N_1} \sim m_{3/2}$ and hence sub-TeV scale RH neutrino may face strong constraints from non-observation of neutralinos and/or charginos.

The existence of a pGF coupled to Higgsino and the charged leptons offers an interesting possibility to search for this fermion at colliders through the displaced vertex signal. It is shown that such a signal can be observed \cite{Lavignac:2020yld} for the parameters of the model required to fit the neutrino data.

\section*{Acknowledgements}
The work of KMP is partially supported by a research grant under INSPIRE Faculty Award (DST/INSPIRE/04/2015/000508) from the Department of Science and Technology, Government of India. The computational work reported in this paper was performed on the High Performance Computing (HPC) resources (Vikram-100 HPC cluster) at the Physical Research Laboratory, Ahmedabad.

\appendix
\section{Sneutrino VEVs and $k$ factor}
\label{AppA:kfactor}
In this Appendix, we show the origin of Eq. (\ref{k}) in the present framework. The procedure is similar to the one carried out previously in \cite{Joshipura:2002fc,Giudice:1992jg}. When the scalar components of $\hat{N}_\alpha$ take VEVs, the bilinear terms in $W$ are given by
\be \label{}
W \supset -\, \epsilon_\alpha\, \hat{L}_\alpha \hat{H}_2\, -\, \mu\, \hat{H}_1 \hat{H}_2\,.
\ee
The $F$ and $D$ term potential evaluated from the above terms is
\beqa \label{}
V_{F,D} &=& \left|\mu \tilde{H}_1 + \epsilon_\alpha \tilde{L}_\alpha \right|^2 + \left(|\mu|^2+|\epsilon_\alpha|^2 \right) |\tilde{H}_2|^2 + \frac{1}{8}(g^2 + g^{\prime 2})\left(|\tilde{H}_1|^2-|\tilde{H}_2|^2 + |\tilde{L}_\alpha|^2 \right)^2,
\eeqa
where index $\alpha$ is summed over. 
Further, the relevant soft terms can be written as
\be \label{}
V_{\rm soft} = m_1^2\, |\tilde{H}_1|^2 + m_2^2\, |\tilde{H}_2|^2 + m_\alpha^2\, |\tilde{L}_\alpha|^2 + \left\{ B_\mu\, \mu\, \tilde{H}_1 \tilde{H}_2 + A_\alpha\, \epsilon_\alpha\, \tilde{L}_\alpha \tilde{H}_2 + {\rm h.c.} \right\}\,.
\ee

The term proportional to $\epsilon_\alpha$ form $W$ can be rotated away with the following redefinitions of the fields
\be \label{Rbasis}
\mu \hat{H}_1+ \epsilon_\alpha \hat{L}_\alpha \to \mu \hat{H}_1\,,~~~~-\epsilon_\alpha \hat{H}_1 + \mu \hat{L}_\alpha \to \mu \hat{L}_\alpha\,.\ee
In this new basis, $V_{F,D}$ does not contain terms involving $\epsilon_\alpha$ while $V_{\rm soft} $ gets replaced by
\be \label{}
V_{\rm soft} \to V_{\rm soft}\,-\, \left\{\frac{\epsilon_\alpha}{\mu} (m_1^2 - m_\alpha^2)\, \tilde{H}_1^\dagger \tilde{L}_\alpha\,+\,\left(B_\mu - A_\alpha\right) \epsilon_\alpha\, \tilde{L}_\alpha \tilde{H}_2\,+{\rm h.c.} \right\}\,.
\ee
In the new basis, the VEV of $\tilde{L}_\alpha$ is denoted by $\omega_\alpha$. It is related to the VEV $\omega^\prime_\alpha$ in the original basis by Eq. (\ref{omega}) as  can be explicitly seen from Eq. (\ref{Rbasis}). Using the definition Eq. (\ref{k}) and minimizing the full potential, we obtain
\be \label{}
k_\alpha \simeq \frac{v_1}{\mu}\frac{(m_1^2 - m_\alpha^2) - \mu \tan\beta \left(B_\mu - A_\alpha\right)}{m_\alpha^2 + \frac{1}{2} M_Z^2\, \cos2\beta}\,.\ee
The first term in $k_\alpha$ can be suppressed for universal soft masses, however the difference $B_\mu - A_\alpha$ can be arbitrary and hence $k_\alpha$ need not  be small. For flavour universal soft terms, $k_\alpha \equiv k$ can be universal at SUSY breaking mediation scale and difference among $k_\alpha$ can arise through renormalization group evolution.

\section{On the size of 1-loop contributions to the neutrino masses}
\label{AppB:loop}
The resulting values of the tree level neutrino mass matrix $m_\nu^{(0)}$ and 1-loop corrected seesaw contribution $\delta m_\nu^{(1)}$ as defined in Eqs. (\ref{mnu0}) and (\ref{dmnu}) are obtained for example, in case of solution BP 4 as
\beqa \label{mnu_ex}
m_\nu^{(0)} & = & \left(
\begin{array}{ccc}
 5.84\, -0.319 i & -8.379+0.932 i & -6.071+0.675 i \\
 -8.379+0.932 i & -21.508-2.723 i & -34.962-1.322 i \\
 -6.071+0.675 i & -34.962-1.322 i & -11.29-1.429 i \\
\end{array}
\right)\,{\rm meV}\,, \nonumber \\
\delta m_\nu^{(1)} & = & \left(
\begin{array}{ccc}
 11.491 & -0.006 & 0.005 \\
 -0.006 & 0.419 & -0.298 \\
 0.005 & -0.298 & 0.219 \\
\end{array}
\right)\, {\rm meV}\,. \eeqa 
The $(1,1)$ element of the neutrino mass matrix receives significant correction from $\delta m_\nu^{(1)}$ while the other elements are dominantly determined by tree level contribution. This characteristic has also been seen in the other benchmark solutions listed in Table \ref{tab:sol}.

In the above, we considered 1-loop self-energy diagram of light neutrinos with Higgs/$Z$ and RH neutrinos in the loop. In general, the neutrino masses also receive corrections from SUSY particles in the loop. A  comprehensive discussion on such corrections is given in \cite{Dedes:2006ni}. We outline these contributions and discuss their magnitudes in the context of our model. 
\begin{enumerate}
\item Neutralino - $\tilde{\nu}_R$ loop: This contribution arises when both Higgs/$Z$ and RH neutrinos are replaced by their SUSY partners in the self-energy diagram. The contribution is $R$-parity conserving and it goes like 
\be \label{dmnu_1}
(\delta m_{\nu})_{\alpha \beta} \sim \frac{1}{16 \pi^2} M_{N}\, \lambda_\alpha \lambda_\beta\, \left(B_0 \left[m_\nu^2, m_{\tilde{\nu}^+_R}^2, M_N^2 \right] - B_0 \left[m_\nu^2, m_{\tilde{\nu}^-_R}^2, M_N^2 \right] \right)\,,
\ee
where $m_{\tilde{\nu}^{\pm}_R}$ are masses of CP even and odd RH sneutrinos. We have used the general result derived in \cite{Dedes:2006ni} to evaluate the above contribution. Apparently,  the above contribution depends on the mass spectrum of sneutrinos and hence its complete determination would require specification of the SUSY breaking mechanism. The contribution vanishes when the RH sneutrinos are degenerate as can be seen from Eq. (\ref{dmnu_1}). 

\item Neutralino-$H/Z$ loop: This arises when RH neutrino is replaced by neutralinos in the loop considered initially. This contribution arises through R-parity violating mixing between neutrino and neutralino and it goes as
\be \label{dmnu_2}
(\delta m_\nu)_{\alpha \beta} \sim \frac{1}{16 \pi^2} M_{1}\,  \left(\frac{\omega_\alpha \omega_\beta}{M_1^2} \right)\, \frac{m_{H/Z}^2}{M_1^2}\, \ln \frac{M_1^2}{m_{H/Z}^2}\,,
\ee
where $\omega_\alpha = k \epsilon_\alpha$ as defined in Eq. (\ref{omega}) and $M_1$ is a common neutralino mass scale. With $M_1 = 1$ TeV and the values of $\omega_\alpha$ obtained for solution BP 2, the largest magnitude of this contribution is ${\cal O}(10^{-6})$ eV. Hence, it is sub-dominant in comparison to loop corrected seesaw contribution given in Eq.  (\ref{mnu_ex}).

\item Neutralino - $\tilde{\nu}_L$ loop: 
This contribution also depends on the details of soft SUSY breaking sector. It vanishes for the degenerate masses of CP even and odd $\tilde{\nu}^\pm_L$. However, in the presence of $R$-parity violating soft terms, $\tilde{\nu}^\pm_L$ mix with the other CP even and odd neutral scalars and the mass splitting between $\tilde{\nu}^\pm_L$ eigenstates gets induced. The contribution to the neutrino mass is approximated as \cite{Dedes:2006ni}
\be \label{dmnu_3}
(\delta m_{\nu})_{\alpha \beta} \sim \frac{1}{16 \pi^2} M_{1}\, g^2\, \left( \frac{M_1}{m_{\tilde{\nu}_L}}\right)^2\, \frac{b_\alpha b_\beta \tan^2\beta}{(M_N^2 - m_{\tilde{\nu}_L}^2)^2}\,,
\ee
where $b_\alpha = B \epsilon_\alpha$. For typical values of the parameters, $M_1 = 1$ TeV, $m_{\tilde{\nu}_L}= 2 $ TeV, $\tan\beta = 2$ and $b_\alpha= 1.0$ GeV$^2$, one obtains $\delta m_\nu \sim {\cal O}(10^{-5})$ eV.

\item Chargino-slepton loop: This contribution is also computed in \cite{Dedes:2006ni} but assuming vanishing VEVs for sneutrinos. We estimate this considering nonzero sneutrino VEVs and find that the dominant contribution is given by
\be \label{dmnu_4}
(\delta m_{\nu})_{\alpha \beta} \sim \frac{1}{16 \pi^2} m_l\,\tan\beta \frac{\omega_\alpha}{M_C}\,\frac{b_\beta}{M_{H^+}^2}\,,
\ee
where $m_l$ is charged lepton mass, $M_2$ is chargino mass. Another similar contribution also arise when $b_\beta$ is replaced by $\epsilon_\beta m_l$ in the above equation. These contributions are suppressed for $\omega_\alpha \ll \epsilon_\alpha$. For typical values,  $M_2 = 1$ TeV, $M_{H^+} = 2$ TeV, $\tan\beta=2$, $b_\alpha= 1.0$ GeV$^2$ and $\omega_\alpha = k \epsilon_\alpha$, the largest contribution from Eq.  (\ref{dmnu_4}) for solution BP 2 is estimated as $\delta m_\nu \sim 10^{-8}$ eV which is again negligibly small compared to the 1-loop seesaw corrected contribution. 
\end{enumerate}

Therefore, the 1-loop correction to the neutrino masses from the SUSY particles in the loop is insignificant in comparison to the loop corrected seesaw contribution that we have taken into consideration. The 2-loop correction to the seesaw is expected to remain smaller than 1-loop contribution as it is further suppressed by factor of $\lambda^2_\alpha/(4 \pi)^2$. Hence, the results and predictions for the various observable parameters that we have derived using $m_\nu$ as given in Eq.  (\ref{mnu}) are robust.

\section{Neutrino spectrum with $\kappa = 0$}
\label{AppC:vanishing_kappa}
In this Appendix, we show the same results as displayed in Tables \ref{tab:sol}, \ref{tab:nudecay} and \ref{tab:clmajoron} but for vanishingly small $\kappa$ and $\tan\beta=1.5$. These cases correspond to vanishingly small mixing between neutralino and RH neutrinos. As it can be seen, one can reproduce neutrino mass spectrum consistent with all the current experimental bounds in this case as well. For BP 6 and BP 7, the predicted value of the sum of light neutrino masses is larger than the current Planck limit. However, the neutrino lifetime obtained in these cases are shorter than the age of the universe as it can be seen from the corresponding values in Table \ref{tab:nudecay_k0}. Hence, these solutions can evade the Planck constraint \cite{Chacko:2019nej} as discussed in section \ref{sec:lmultau}.
\begin{table}[!h]
\begin{center}
\begin{tabular}{ccccc} 
\hline
\hline
 ~~~~~~Parameters ~~~~~~& ~~~~~~BP 5~~~~~~  & ~~~~~~BP 6~~~~~~ & ~~~~~~BP 7~~~~~~ & ~~~~~~BP 8~~~~~~ \\
\hline
$U$ [GeV] & $10^3$ & $10^4$  & $10^5$ &  $10^6$ \\
$\kappa$ & $0$ & $0$  & $0$ &  $0$ \\
$\lambda$ &  $-0.1+0.0004\, i$ & $0.1168-0.0013\, i$ & $-0.9998 - 0.3\, i$ & $0.2973 + 0.0891\, i$ \\
$\epsilon$ & 0.0139 & -0.0377 & -0.0015 & -0.0007\\
$\phi$ & -0.7764 & 2.1609 & -2.1093 & 1.7723\\
$\lambda_\tau$ & $7.59 \times 10^{-8}$ & $5.031 \times 10^{-7}$ & $4.261 \times 10^{-7}$  & $9.032 \times 10^{-7}$\\
$r_\mu$ & 1.0749 & 0.7156 & -0.8108 & 0.2771\\
$r_e$ & -43.4421 & -11.7682 & 85.9164 & 64.9009\\
$k$ & 4.2802 & 0.082 & 0.0118 & 0.0014 \\
\hline
$\Delta m^2_{21}$ [$10^{-5}$ eV$^2$] & $7.43$ & 7.43 & 7.42 & 7.42\\
$\Delta m^2_{31}$ [$10^{-3}$ eV$^2$] & $2.513$ & 2.515  & 2.517  & 2.517\\
$\sin^2 \theta_{12}$ & 0.309 & 0.303 & 0.304 & 0.304\\
$\sin^2 \theta_{23}$ & 0.6 & 0.596 & 0.573 & 0.573 \\
$\sin^2 \theta_{13}$ & 0.0221 & 0.0221 & 0.0222 & 0.0222\\
$\sin \delta_{\rm CP}$ & -0.14 & -0.42 & -0.29 & -0.29\\
\hline
$\sum m_{\nu_i}$ [eV] & 0.252 & 0.23 & 0.093 & 0.093\\
$|m_{\beta \beta}|$  [eV] & 0.079 & 0.071 & 0.02 & 0.02\\
$M_{N_1}$ [GeV] & 1.39 & 40.7 & 133.4  & 91.2\\
$M_{N_2}$ [GeV] & 99.31 & 1147  & 104317 & 310266\\
$M_{N_3}$ [GeV] & 100.7 & 1188  & 104451 & 310357\\
\hline
\hline
\end{tabular}
\end{center}
\caption{Same as Table \ref{tab:sol} but with $\kappa = 0$.}
\label{tab:sol_k0}
\end{table}
\begin{table}[h!]
\begin{center}
\begin{tabular}{ccccc} 
\hline
\hline
~~~Parameters ~~~& ~~~~~~BP 5~~~~~~  & ~~~~~~BP 6~~~~~~ & ~~~~~~BP 7~~~~~~ & ~~~~~~BP 8~~~~~~ \\
\hline
$|\eta_0|$ & 0.341 & 0.366  & 2.497 &  2.495 \\
${\rm Max.}\{|g_{ii}|\}$ & $1.5 \times 10^{-14}$ & $1.4 \times 10^{-15}$ & $1.2 \times 10^{-16}$ & $1.2 \times 10^{-17}$\\
$|g_{32}|$ &  $5.9 \times 10^{-14}$ &$5.9 \times 10^{-15}$ & $1.4 \times 10^{-17}$ & $1.4 \times 10^{-18}$ \\
$|g_{31}|$ &  $6.7 \times 10^{-14}$ & $6.0 \times 10^{-15}$ & $3.1 \times 10^{-18}$ & $3.1 \times 10^{-19}$ \\
$\tau_\nu$ [sec.] & $9.1 \times 10^{13}$ & $9.9 \times 10^{15}$ & $3.2 \times 10^{21}$ & $3.2 \times 10^{23}$\\
\hline
\hline
\end{tabular}
\end{center}
\caption{Same as Table \ref{tab:nudecay} but with $\kappa = 0$.}
\label{tab:nudecay_k0}
\end{table}
\begin{table}[h!]
\begin{center}
\begin{tabular}{ccccc} 
\hline
\hline
~~~Parameters ~~~& ~~~~~~BP 5~~~~~~  & ~~~~~~BP 6~~~~~~ & ~~~~~~BP 7~~~~~~ & ~~~~~~BP 8~~~~~~ \\
\hline
${\rm Max.}\{|h_{\alpha \beta}|\}$ &  $10^{-16}$ & $10^{-15}$ & $10^{-14}$ & $10^{-13}$ \\
\hline
\hline
\end{tabular}
\end{center}
\caption{Same as Table \ref{tab:clmajoron} but with $\kappa = 0$.}
\label{tab:clmajoron_k0}
\end{table}

\newpage

\bibliography{references}
\end{document}